\documentclass[utf8]{FrontiersinVancouver} 

\usepackage{url,hyperref,lineno,microtype,subcaption}

\usepackage[onehalfspacing]{setspace}

\def\keyFont{\fontsize{8}{11}\helveticabold }
\def\firstAuthorLast{Torun {et~al.}} 
\def\Authors{Tugba Torun, Ameer Taweel, and Didem Unat\,$^{*}$}

\def\Address{
ParCoreLab, Koç University, Computer Science and Engineering, Istanbul, Türkiye}

\def\corrAuthor{Didem Unat}
\def\corrEmail{dunat@ku.edu.tr}

\usepackage{algorithm}
\usepackage{algpseudocode}
\usepackage{multirow}
\usepackage{float}
\usepackage{booktabs}

\newcommand{\fy}{\mathcal{F}}
\newcommand{\sy}{\mathcal{S}}
\newcommand{\xy}{\mathcal{X}}
\newcommand{\gentensor}[0]{\mbox{\textsc{GenTensor}}}
\newcommand{\featensor}[0]{\mbox{\textsc{FeaTensor}}}

\begin{document}
\onecolumn
\firstpage{1}

\title[A Sparse Tensor Generator with Efficient Feature Extraction]{A Sparse Tensor Generator with Efficient Feature Extraction} 

\author[\firstAuthorLast ]{\Authors} 
\address{} 
\correspondance{} 

\extraAuth{}

\maketitle

\begin{abstract}

\section{}

Sparse tensor operations are increasingly important in diverse applications such as social networks, deep learning, diagnosis, crime, and review analysis. 
However, a major obstacle in sparse tensor research is the lack of large-scale sparse tensor datasets.
Another challenge lies in analyzing sparse tensor features, which are essential not only for understanding the nonzero pattern but also for selecting the most suitable storage format, decomposition algorithm, and reordering methods. However, due to the large size of real-world tensors, even extracting these features can be computationally expensive without careful optimization.
To address these limitations, we have developed a smart sparse tensor generator that replicates key characteristics of real sparse tensors. Additionally, we propose efficient methods for extracting a comprehensive set of sparse tensor features. The effectiveness of our generator is validated through the quality of extracted features and the performance of decomposition on the generated tensors.
Both the sparse tensor feature extractor and the tensor generator are open source with all the artifacts available at \url{https://github.com/sparcityeu/FeaTensor} and \url{https://github.com/sparcityeu/GenTensor}, respectively.

\tiny
 \keyFont{ \section{Keywords:} sparse tensor, tensor generators, feature extraction, synthetic data generation, shared memory parallelism} 

\end{abstract}


\section{Introduction}
\label{sec:intro}

Several applications such as social networks, deep learning, diagnosis, crime, and review analysis require the data to be processed in the form of multi-dimensional arrays, namely \textit{tensors} \citep{kuang2014tensor, leclercq2018tensor, acar2007multiway, mu2011empirical, jukic2013noninvasive, yelundur2019detection}.
Tensors are extensions of matrices that can have three or more dimensions, or so-called \textit{modes}.
Tensor decomposition techniques such as CPD (Canonical Polyadic decomposition) and Tucker are widely used to analyze and reveal the latent features of such real-world data in the form of sparse tensors \citep{kolda2009tensor, goulart2015tensor}.
For making the tensor decomposition methods faster or more memory-efficient, numerous storage formats and partitioning or reordering schemes are introduced in the literature \citep{li2019efficient, acer2018improving, kaya2018parallel, li2018hicoo, shivakumar2021efficient, helal2021alto}.
However, the performance of these schemes highly depends on the sparsity pattern of the tensor.

Efficiently extracting the sparsity pattern of a given tensor is essential for optimizing various aspects of tensor analysis and manipulation. The structural features can inform format selection, aiding in determining the most suitable storage format. It can influence the algorithm selection, with different tensor operations exhibiting varying performance based on the sparsity characteristics. Understanding the sparsity pattern also provides insights into the performance of decomposition techniques. 
In a recent work, \citep{sun2021input}, some tensor features 
are used to automatically predict the best storage format for a sparse tensor via machine learning techniques. However, the work considers the features for only a single mode, which may cause lack of some critical intuition from other dimensions.
 
Sparse tensors in real-world scenarios often exhibit extreme sparsity, with densities as low as $10^{-15}$.
Unlike the sparse matrices often containing at least one nonzero in their rows and columns, sparse tensors contain numerous empty {\em fibers} and {\em slices}, which are one- and two-dimensional fragments of tensors. A naive approach for extracting sparse tensor features involves traversing the tensor nonzeros in coordinate (COO) format and updating the nonzeros of respective slices and fibers. However, this approach becomes impractical for large tensors with increasing dimensions.
Another approach \citep{sun2021input} is to assume that the tensor is already sorted to extract the features, yet it reveals the features from only a single mode order. 
Since the sizes of a sparse tensor diverge a lot along different modes,
focusing the features solely on one perspective might lead to the loss of crucial structural information inherent in the tensor.

To tackle these challenges, we develop a sparse tensor feature extraction framework, \featensor{}, which extracts a detailed and exclusive set of sparse tensor features, encapsulating the features along all modes.
It extends the feature set by including important size-independent features such as coefficient of variation and imbalance to gain a deeper insight into the nonzero distribution. 
Additionally, \featensor{} offers four alternative feature extraction methods for efficiency, providing flexibility to select the most suitable method based on machine and tensor characteristics.

By utilizing the generated features, machine learning tools can be used to reveal the most suitable storage format, partitioning, or reordering method that fits best with that tensor. 
Nevertheless, the primary challenge facing this research stems from the necessity of having thousands of samples to train machine learning models.
Conversely, the majority of multi-dimensional real-world data require manual cleaning to become readily usable in research. 
Meanwhile, efforts to gather publicly available real-world data as sparse tensor collections yield only a few instances \citep{jeon2015haten2, smith2017frostt}. 
Moreover, tensor sizes can be large, making it inconvenient to download, read, and use them in computation. A tensor generator for performance analysis purposes can be handy, enabling the study of algorithms by generating tensors on the fly and discarding them if necessary.

To address these gaps in the literature, we propose a smart sparse tensor generator, \gentensor{}, which accounts for key tensor features during generation. This generator produces sparse tensors with characteristics closely resembling those of real-world tensors, enabling the creation of large-scale sparse tensor datasets.
A key advantage of \gentensor{} is its use of size-independent features -such as the coefficient of variation, imbalance, and density- allowing flexible generation of tensors at different scales. This flexibility enables researchers to efficiently prototype and benchmark tensor algorithms on smaller, representative tensors before scaling to real-world datasets.
The effectiveness of \gentensor{} is validated by comparing the extracted features and CPD performance of the generated tensors against both naïve random tensors and real-world tensors. These features are obtained using our dedicated feature extraction tool, \featensor{}, which ensures accurate characterization and comparison of tensor properties.

The main contributions of this work are:
\begin{itemize}
    \item A sparse tensor feature extraction framework, \featensor{}, is developed. It includes four different feature extraction methods, which can be used alternatively depending on the computation needs and the characteristics of input tensors. All methods in \featensor{} are parallelized using OpenMP.

    \item We develop a smart sparse tensor generator, \gentensor{}, which considers significant tensor features. It can be used to generate an artificial tensor dataset in which the properties and characteristics of the generated tensors are similar to the real-world tensors. \gentensor{}  is parallelized with OpenMP for faster generation.

    \item Experiments on several sparse tensors validate that all feature extraction methods in $\featensor{}$ give the exact results as feature sets. We present the runtime comparison of the methods in $\featensor{}$ to guide users to exploit the one that is most appropriate to their needs.

    \item We demonstrate the effectiveness of the proposed tensor generator  \gentensor{} in terms of feature quality, tensor decomposition performance, and sensitivity to seed selection.

    \item Both tools are open source \footnote{\url{https://github.com/sparcityeu/FeaTensor}}
    \footnote{\url{https://github.com/sparcityeu/GenTensor}}, accompanied by comprehensive documentation and illustrative examples tailored for the community's usage.
\end{itemize}

This manuscript is organized as follows.
Section \ref{sec:background} provides the background information on sparse tensors and tensor decomposition.
The proposed feature extraction tool is introduced in Section \ref{sec:featensor}.
In Section \ref{sec:generator}, we present our sparse tensor generator, \gentensor{}.
Experimental results are reported in Section \ref{sec:evaluation}.
We discuss the related works in Section \ref{sec:related-work} and conclude in Section \ref{sec:conclusion}.


\section{Materials and Methods}

\subsection{Background}
\label{sec:background}

\begin{figure}[!tbp]
    \begin{center}
    \includegraphics[width=0.65\linewidth]{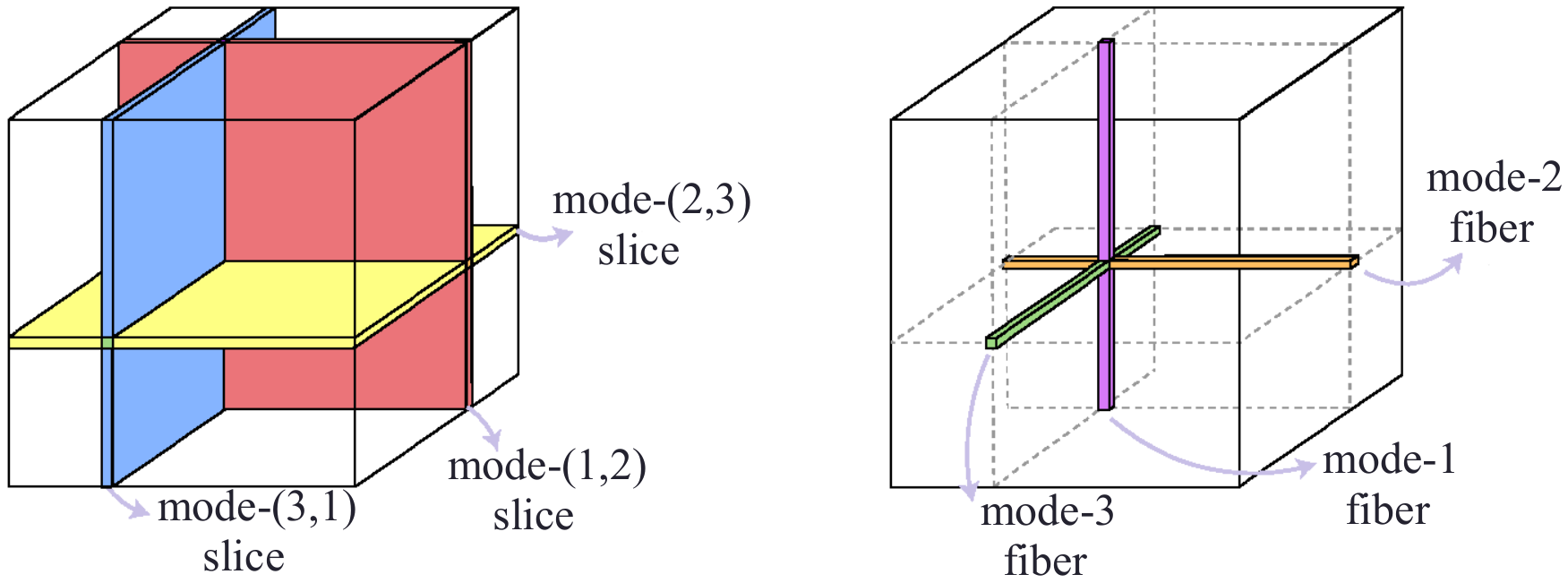}
    \caption{Sample slice and fibers of a 3-mode tensor.}
    \label{fig:slice-fiber}
    \end{center}
\end{figure}

A tensor with $M$ dimensions is called an $M$-$mode$ or $M$th $order$ 
tensor, where the mode count, $M$, is referred to as the $order$ of tensor.
Mode $m$ of a tensor refers to its $m$th dimension.
\textit{Fibers} are defined as the one-dimensional sections of a tensor obtained by fixing all but one index.
\textit{Slices} are two-dimensional sections of a tensor obtained by fixing every index but two.
Figure \ref{fig:slice-fiber} depicts sample slice and fibers of a 3-mode tensor. 
The numbers in the naming of slices and fibers derive from the mode indices that are not fixed while forming them.
For instance, $\xy(i,:,:)$ is a mode-(2,3) slice, and $\xy(i,j,:)$ is a mode-3 fiber of a 3-mode tensor $\xy$, where a tensor element with indices $i, j, k$ is denoted by $\xy(i,j,k)$.
A fiber (slice) is said to be a $nonzero$ $fiber$ ($slice$) if it contains at least one nonzero element; and an $empty$ $fiber$ ($slice$), otherwise.



To reveal the relationship of data across different modes, tensor decomposition techniques are widely used.
Canonical Polyadic Decomposition (CPD) and Tucker decomposition are the two most popular ones among them.
In Tucker decomposition, a tensor is decomposed into a much smaller core tensor and a set of matrices; whereas in CPD, a tensor is factorized as a set of rank-1 tensors, which can be considered as a generalization of matrix singular value decomposition (SVD) method for tensors.

There are various implementations for CPD and each has different storage schemes and reordering algorithms proposed specifically for that scheme.
SPLATT~\citep{smith2015splatt} and ParTI~\citep{li2018parti} are two commonly-used libraries for CPD.
ParTI is a parallel tensor infrastructure that supports essential sparse tensor operations and tensor decompositions on multicore CPU and GPU architectures. 
SPLATT provides a shared-memory implementation for CPD while adopting a medium-grain partitioning for sparse tensors for parallel execution of CPD.


\subsection{Sparse Tensor Feature Extraction}
\label{sec:featensor}

The features of a sparse tensor have the capacity to reveal the most suitable storage format, partitioning or reordering method that fits best with that tensor if well-examined.
The main challenge in extracting the sparse tensor features is to determine which and how many fibers and slices are nonzero.
This is because the real-world sparse tensors are highly sparse and contain many empty fibers and slices.
One naive approach for sparse tensor feature extraction is to traverse the tensor nonzeros in coordinate (COO) format and update the number of nonzeros of the respective slices and fibers. 
However, this approach is not practical for large tensors since the real-world tensors often have huge numbers of fibers and slices.
For instance, some real tensors have more than $10^{19}$ fibers, and even storing a boolean array of such a large size requires a space of 10 million Terabytes, which is impractical in modern machines (See the properties of some real tensors in Table \ref{table:dataset}).
To overcome these challenges, 
we propose and implement four alternative methods for sparse tensor feature extraction. 

\subsubsection{Feature Set}
We consider three \textit{kinds} of statistics for sparse tensors: (i) nonzeros per nonzero slices, (ii) nonzeros per nonzero fibers, (iii) nonzero fibers per nonzero slices. 
Table \ref{table:tensor_feat_global} depicts the global features of the tensor that are independent of these kinds.
Table \ref{table:tensor_feat_mode} shows the mode- and kind-dependent features that we have extracted for all of these three kinds and for all modes.
Here, by referring to all modes, we mean all possible angles that a slice or fiber can have through fixing different modes. 
For example, mode-(1,2), mode-(2,3), and mode-(3,1) slices represent all modes for slices of a 3-mode tensor. 
For ease of expression, we refer to the set of slices (fibers) along different modes as $slice$-$modes$ ($fiber$-$modes$).

A tensor $\xy$ of size $I_1 \times I_2 \times  \dots \times I_M$ is assumed to be given as an input in Coordinate format, i.e. the extension of matrix-market format for tensors.
In the formulas, \texttt{nz}($\cdot$) is used as a function returning the number of nonzeros.
$\fy_m$ refers to the set of fibers in mode-$m$, whereas $\sy_{k,\ell}$  refers to the set of slices obtained by fixing indices in modes $k$ and $\ell$.
In the feature names, $\bullet_{nz}$ refers to considering only nonzero kind-entries in the computation, i.e. nonzero fibers or slices, by ignoring the empty fibers or slices. $\bullet_{all}$ refers to considering all kind-entries in the computation, including the empty ones.
For instance, for the nonzeros-per-fiber kind in mode 1, $n_{all}$ gives the number of all mode-1 fibers including empty, whereas $n_{nz}$ gives the number of only nonzero fibers along mode-1.



\begin{table}[!tbp]
\begin{minipage}[t]{0.45\linewidth}
\caption{Extracted global features of tensors. }
\label{table:tensor_feat_global}
\resizebox{0.95\columnwidth}{!}{%
\begin{tabular}{lll}
\toprule
Feature   & Description & Formula                                      \\ 
\midrule
size\_m                  & Size of tensor in mode m & $I_m$                           \\
nnz                      & Number of nonzeros  & $\texttt{nz}(\xy)$                                    \\
$\mathrm{d}_\mathrm{nz}$                 & Density of nonzeros  & $\texttt{nz}(\xy)/ \prod I_m$                           \\
$\mathrm{nfib}_\mathrm{all}$                 & Number of all fibers  &  $\sum | \fy_m |$    \\
$\mathrm{nslc}_\mathrm{all}$                  & Number of all slices    & $\sum | \sy_{k,\ell} |$ \\
$\mathrm{nfib}_\mathrm{nz}$                 & Number of nz fibers  & $\sum | \texttt{nz} (\fy_m)|$    \\
$\mathrm{nslc}_\mathrm{nz}$                  & Number of nz slices    & $\sum | \texttt{nz} (\sy_{k,\ell})|$ \\
$\mathrm{d}_\mathrm{fib}$ & Density of nz fibers & $\mathrm{nfib}_\mathrm{nz} / \mathrm{nfib}_\mathrm{all}$    \\
$\mathrm{d}_\mathrm{slc}$ & Density of nz slices  & $\mathrm{nslc}_\mathrm{nz} / \mathrm{nslc}_\mathrm{all}$ \\
\bottomrule
\end{tabular}
}
\end{minipage}
\hfill
\begin{minipage}[t]{0.53\linewidth}
\caption{Mode- and kind-dependent features that are extracted for each mode and for each kind: (i) nonzeros per slice, (ii) nonzeros per fiber, (iii) fibers per slice.  }
\label{table:tensor_feat_mode}
\resizebox{\columnwidth}{!}{%
\begin{tabular}{lll}
\toprule
Feature          & Description     & Formula             \\
\midrule
$\mathrm{n}_\mathrm{all}$      & All count including empty    &    *     \\
$\mathrm{n}_\mathrm{nz}$       & Nonzero count    &  *      \\
nz\_density   & Nonzero sparsity   & $\mathrm{n}_\mathrm{nz} / \mathrm{n}_\mathrm{all}$       \\
max           & Maximum value   &  *     \\
min           & Minimum value    &  *    \\
dev           & Deviation  & $\text{max} - \text{min}$      \\
sum           & Summation of values     &  *   \\
$\mathrm{avg}_\mathrm{all}$           & Average value  & sum /$\mathrm{n}_\mathrm{all}$   \\
$\mathrm{imbal}_\mathrm{all}$         & Imbalance & $(\text{max} -\mathrm{avg}_\mathrm{all})/\text{max}$\\
$\mathrm{stDev}_\mathrm{all}$         & Standard deviation   &  *    \\
$\mathrm{cv}_\mathrm{all}$            & Coefficient of variation & $\mathrm{stDev}_\mathrm{all} / \mathrm{avg}_\mathrm{all}$  \\
$\mathrm{avg}_\mathrm{nz}$   & Average by excluding empty & sum /$\mathrm{n}_\mathrm{nz}$  \\
$\mathrm{imbal}_\mathrm{nz}$ & Imbalance by excluding empty & $(\text{max} -\mathrm{avg}_\mathrm{nz})/\text{max}$ \\
$\mathrm{stDev}_\mathrm{nz}$ & stDev by excluding empty  &  * \\
$\mathrm{cv}_\mathrm{nz}$    & cv by excluding empty & $\mathrm{stDev}_\mathrm{nz} / \mathrm{avg}_\mathrm{nz}$  \\
\bottomrule
\end{tabular}
}
\footnotesize{*Omitted because the formula is too complex or trivial.}
\end{minipage}
\end{table}

In addition to the tensor features utilized in the related work \citep{sun2021input}, we also include the features of load imbalance,  standard deviation, and coefficient of variation.
In \citep{nisa2019load}, it is shown that a high standard deviation of fiber length causes inter-warp load imbalance and low occupancy; whereas high standard deviation of the slice volume is related to significant inter-thread-block load imbalance.
The coefficient of variation is another commonly-used metric for analysis which allows to compare between data sets with widely different means.
For example, it is used to evaluate the dispersion of the number of nonzero elements per row for sparse matrix computations~\citep{abu2013auto}.

\subsubsection{Extraction Workflow}
The feature extraction process consists of two main stages, namely \textit{array construction} and the \textit{final reduction} phases.
In the array construction phase, we construct three arrays for all modes: (i) $n^{nz}_{slc}$ (number of nonzeros per slice), (ii) $n^{nz}_{fib}$ (number of nonzeros per fiber), (iii) $n^{fib}_{slc}$ (number of fibers per slice). 
Here and hereafter, only the nonzero slices and nonzero fibers are considered when referring to slices and fibers, if not stated otherwise.
Then in the final reduction phase, we extract all the sparse tensor features by traversing or applying a reduction on these smaller arrays.
Both the array construction and the final reduction phases of \featensor{} are parallelized using OpenMP.

The arrays of $n^{nz}_{slc}$ and $n^{fib}_{slc}$ are constructed for each slice-mode, so that there will be ${M \choose 2}$ of them.
The array $n^{nz}_{fib}$ is constructed for each fiber-mode, hence yielding $M$ different $n^{nz}_{fib}$ arrays.
For instance, for a 3-mode tensor, there are 3 different types of statistics for $n^{nz}_{slc}$ and $n^{fib}_{slc}$ which correspond to mode-(1,2), (2,3), and (3,1) slices; whereas for $n^{nz}_{fib}$, there are 3 different types of statistics corresponding to mode-1, 2, and 3 fibers.
Since there are 11 global features and 15 mode- and kind-dependent features, the total number of features extracted for a 3-mode tensor is $15 \times 3 \times 3 + 11 = 146$ in total.
For a 4-mode tensor, slices are along ${4 \choose 2}=6$ different mode pairs, and fibers are along 4 different modes, so in total $15\times 6\times 2 +15\times4+11=251$ features are considered.
For 5-mode tensors, slices are along ${5 \choose 2}=10$ different mode pairs, and fibers are along 5 different modes, yielding a total of $15\times 10 \times 2 +15\times 5+11=386$ features.

One approach in \featensor{} is to consider the slice and fiber modes in relation to mode execution orders.
By saying $mode$-$order$ $\langle i,j,k \rangle$, we can think that the COO-based index arrays of tensor $\mathcal{X}$ are virtually rearranged to 
$\langle\xy_i,\xy_j,\xy_k\rangle$ instead of the original order $\langle\xy_1,\xy_2,\xy_3\rangle$.
In this context, we consider the slices are obtained by fixing all indices except the last two indices, and the fibers are obtained by fixing all indices except the last one.
For instance, in mode order $\langle 1,2,3 \rangle$, we consider mode-(2,3) slices and mode-3 fibers.
The advantage of this approach is to combine slices and fibers so that their corresponding computations may overlap.

\begin{figure*}[!tbp]
    \begin{center}
    \includegraphics[width=0.87\linewidth]{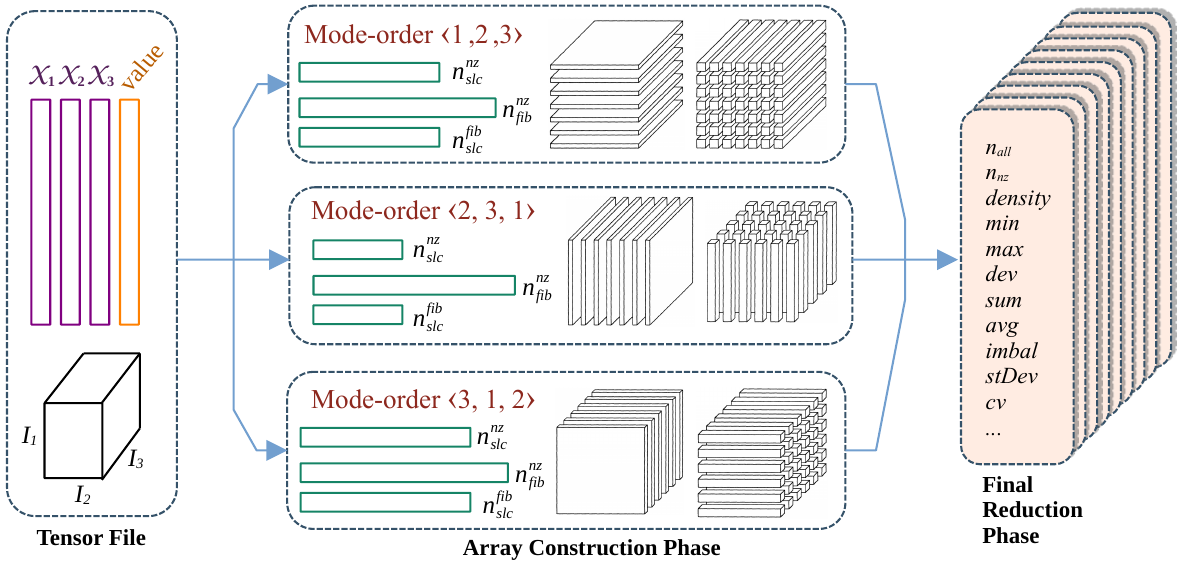}
    \caption{The workflow of feature extraction for a 3-mode tensor using the mode-order approach.}
    \label{fig:workflow}
    \end{center}
\end{figure*}

Note that especially when computing $n^{fib}_{slc}$ arrays, two different fiber modes can be associated with each slice mode.
To be more specific, to extract the number of fibers per mode-$(i,j)$ slices, it is possible to consider either mode-$i$ or mode-$j$ fibers.
For this reason, to cover all slice and fiber modes, one can use both the mode-order set ${\langle 1,2,3 \rangle, \langle 2,3,1 \rangle, \langle 3,1,2 \rangle}$ or the mode-order set ${\langle 1,3,2 \rangle, \langle 2,1,3 \rangle, \langle 3,2,1 \rangle}$.
Figure \ref{fig:workflow} depicts the workflow of the feature extraction process for a 3-mode sparse tensor for the first case.
In the final reduction phase, nine different sets of mode- and kind-dependent features are extracted in parallel, corresponding to three different versions of $n^{nz}_{slc}$, $n^{nz}_{fib}$, and $n^{fib}_{slc}$ arrays from distinct mode-orders.



The mode-order approach is especially practical for 3-mode tensors.
It is because the number of slice-modes and fiber-modes are both equal to three, making a one-to-one mapping possible between slice- and fiber-modes, hence execution on three distinct mode-orders is sufficient to cover all cases.
However, for higher dimensions, it can cause some calculations for $n^{nz}_{fib}$ to be repeated since the number of slice-modes becomes larger than the number of fiber-modes.
Therefore, we obey different approaches in different methods of \featensor{}.

\subsubsection{Extraction Methods}
In order to overcome the memory and speed limitations in trivial feature extraction approaches, we propose four different feature extraction methods that can be used alternatively depending on the computation needs and the characteristics of input tensors.
These methods are heap-based, sorting-based, grouping-based, which is first proposed in this work, and a hybrid method, which is a combination of sorting- and grouping-based methods.

\paragraph{Hash-based method:}
We implement a hash-table-based method to solve the memory issue of the COO-based naive method for large sparse tensors. 
Instead of long arrays that mostly contain zero values, we use a hash table that has keys as slice or fiber indices and values as the number of nonzeros in the corresponding slice or fiber.
Our hashing method excludes ignored dimensions per mode and applies bitwise mixing to ensure a well-distributed hash. 
The hash table is initialized with 100 buckets, though it dynamically resizes as needed.
One main difference of the hash-based method from other methods in \featensor{} is that it does not follow the mode-order approach in Figure \ref{fig:workflow}, i.e. it does not pair and handle some slice and fibers together, but rather extracts the slice features and fiber features independently. 

\paragraph{Sorting-based method:}
Another approach for feature extraction is based on sorting the tensor.
It is the conventional approach for extracting the features or constructing compressed storage formats for tensors in the literature \citep{sun2021input, smith2015tensor, li2018hicoo}. 
It naturally follows the mode-order approach by sorting the tensor indices according to a given mode-order.
After sorting, the array construction phase becomes easier since the nonzeros belonging to the same fiber or slice are positioned consecutively.
The related work \citep{sun2021input} assumes the tensor to be given as sorted and computes the features only for a single mode-order. However, this assumption is not valid when it comes to executing in the upcoming mode-orders.
In other words, a tensor $\mathcal{X}$ sorted in $\langle1,2,3\rangle$ order has to be re-sorted for $\langle2,3,1\rangle$ and $\langle3,1,2\rangle$ orders to execute the features for other modes.
Instead, we find the tensor features along all modes by sorting the related indices of tensor.

\paragraph{Grouping-based method:}
Instead of sorting the tensor nonzeros fully, we group the slices and fibers according to their indices and keep the last indices of the COO format in their original order. 
We first find the number of nonzero fibers and slices by traversing the tensor entries.
We store the indices of these nonzero slices and fibers in a compressed manner and track their nonzero counts in the respective arrays.
This algorithm takes inspiration from the construction process of Compressed Sparse Fiber (CSF) format, which is a generalization of Compressed Sparse Row and Compressed Sparse Column (CSR and CSC) formats for higher dimensions.

Algorithm \ref{alg:feature_extraction} shows the pseudo-code of the grouping-based feature extraction method for the mode-order $\langle1,2,3\rangle$ of a 3-mode tensor. 
In this case, the algorithm considers mode-(2,3) slices and mode-3 fibers.
$cnt1$ is a temporary array that keeps the nonzero count information for all slices (Lines 1-4).
It is used to obtain the number of nonzero slices ($nslc$), the array of nonzero counts for each nonzero slice ($n^{nz}_{slc}$), the array of indices for nonzero slices ($ind_{slc}$), and the array of starting locations for nonzero slices (($xn^{nz}_{slc}$) that is similar to the $row\_ptr$ array in CSR format for matrices (Lines 5-8).
$loc$ and $order$ are temporary arrays that are constructed (Lines 10-16) to obtain the $cnt2$ array, which keeps the nonzero count information for all fibers in each nonzero slice (Lines 18-21).
Finally, $cnt2$ is used to obtain the number of nonzero fibers  ($n^{fib}_{slc}$) and the array of nonzero counts per nonzero fiber ($n^{nz}_{fib}$) in each nonzero slice (Lines 22-23).

\begin{algorithm}[!t]

\begin{minipage}{.99\linewidth}
\caption{Grouping-Based Extraction Method in \featensor{}}
\label{alg:feature_extraction}
\vspace{0.3em}
\hspace*{\algorithmicindent} 
\hspace{-0.3em}
\textbf{Input:} 3-mode tensor with COO-arrays $\mathcal{X}_1$, $\mathcal{X}_2$, $\mathcal{X}_3$ \\
\vspace{0.3em}
\hspace*{\algorithmicindent} 
\hspace{-0.5em}
\textbf{Output:} $n^{nz}_{slc}$, 
 $n^{nz}_{fib}$, $n^{fib}_{slc}$ arrays for mode-order $\langle 1, 2, 3\rangle$ 

\begin{algorithmic}[1]
\State Create the array $cnt1$ of size $I_1$ and initialize to zero
\For {$i \leftarrow 0$ to NNZ}
\State $cnt1 ( \mathcal{X}_1 (i) )  ) +\!+$ \hfill $\rhd$ Temporary array for nonzeros per slices
\EndFor
\State $nslc \leftarrow$ \# nz values in $cnt1$ \hfill $\rhd$ Number of nonzero slices
\State $n^{nz}_{slc} \leftarrow$ nz values in $cnt1$ (size $I_1$)  \hfill $\rhd$ Number of nonzeros per nonzero slice
\vspace{0.3em}
\State $ind_{slc} \leftarrow$ indices of nz values in $cnt1$ (size $I_1$)
\vspace{0.3em}
\State $xn^{nz}_{slc}\leftarrow \texttt{PrefixSum} ( n^{nz}_{slc} )$
\State Create the array $loc$ of size $ind_{slc} (I_1)\!+\!1$
\For {$i \leftarrow 0$ to $nslc$}
\State $loc (ind^{nz}_{slc} (i )) \leftarrow xn^{nz}_{slc}(i)$
\EndFor
\State Create the array $order$ of size $NNZ$
\For {$i \leftarrow 0$ to NNZ}
\State $order ( loc (\mathcal{X}_1 (k))+\!+) \leftarrow k$
\EndFor
\vspace{0.5em}
\For{$i \leftarrow 0$ to $nslc$}
\State Create the array $cnt2$ of size $I_2$ and initialize to zero \hfill $\rhd$ Temporary array for nonzeros per fiber
\For{$j\leftarrow xn^{nz}_{slc}(i)$ to $xn^{nz}_{slc}(i+1)$}
\vspace{0.3em}
\State $cnt2 ( \mathcal{X}_2 (order (j))) +\!+ $
\EndFor
\State $n^{fib}_{slc}(i) \leftarrow$ \# nz in $cnt2$ \hfill $\rhd$ Number of nonzero fibers per nonzero slice
\vspace{0.3em}
\State $n^{nz}_{fib}(i) \leftarrow$ nz values in $cnt2$ (size $n^{fib}_{slc}(i)$) \hfill $\rhd$ Number of nonzeros per nonzero fiber
\EndFor

\end{algorithmic}
\end{minipage}

\end{algorithm}

The memory requirement of this algorithm is $I_1 \!+\! I_2 \!+\! NNZ$ for executing the $(i_1,i_2,:)$ fibers of an $I_1 \!\times\! I_2 \!\times\! I_3$ tensor with $NNZ$ nonzeros. The worst case serial runtime is $\mathcal{O}(I_1+I_2 \mathcal{S}_1)$ where $\mathcal{S}_1$ is the number of nonzero slices in mode 1.
For the general case, $size(m)\times size(m+1)$ is a loose upper bound for mode $m$.

\paragraph{Hybrid method:}
We propose a hybrid method by combining the sorting- and grouping-based methods.
The idea is to utilize different methods for extracting features in different modes within the same tensor. 
This is because the performance of different methods varies depending on the size of the respective mode.
Since we observe that sorting- and grouping-based methods are the best-performing ones at the mode level, we use them interchangeably according to the respective mode size.
For mode $m$, we apply the grouping-based method if $size(m)\times size(m+1) < \lambda$ holds, and we apply the sorting-based method otherwise; 
where $\lambda$ is a predetermined threshold.
The reasoning behind this choice is to limit the expected worst-case runtime for the respective modes when employing 
the grouping-based method.
We set the threshold $\lambda$ as $10^{11}$ based on empirical evaluations, whose details will be discussed in Section \ref{sec:feat-reasult}.

\subsubsection{Extension to Higher Orders}

In \featensor{}, we include two different options for feature extraction of higher-order tensors, namely $all$-$modes$ and $only$-3-$mode$ options.
The $all$-$modes$ option extracts all features along all modes.
In the $only$-3-$mode$ option, we extract the features along only the modes with the three largest sizes.
That is, tensor modes are sorted according to their sizes, the three largest ones are selected, and the features corresponding solely to those modes are extracted.

As expected, the $all$-$modes$ option is significantly more time- and resource-intensive. As the tensor order increases, the total number of features across all modes grows dramatically. The large number of features required for extracting higher-order tensors complicates the evaluation process.
Instead, the $only$-3-$mode$ option is sufficient in most cases. For instance, in our tensor generator, \gentensor{}, features for slices and fibers are required only along specific modes. Therefore, the $only$-3-$mode$ option is adequate for generating tensors with \gentensor{}.

Recall that the mode-order approach used by the sorting-based, grouping-based, and hybrid methods is particularly effective for extracting features along three modes, as the gap between the number of slice modes and fiber modes increases with higher tensor orders.
Therefore, we include the $all$-$modes$ option for higher-order tensors only in the hash-based method, as it does not follow the mode-order approach.

For the $only$-3-$mode$ option, we extract features along three modes of an $M$-mode tensor by treating the indices of these three modes as a 3-mode tensor.
Therefore, this approach aligns with the mode-order approach. As a result, the $only$-3-$mode$ option is available for all methods in \featensor{}.


\subsection{Sparse Tensor Generator}
\label{sec:generator}

Despite the increased need for research in sparse tensors, publicly available sparse tensor datasets remain limited, comprising only a few instances \citep{jeon2015haten2, smith2017frostt}. To address this scarcity and facilitate the study of machine learning models with a larger variety of sparse tensors, as well as to expand the size of open datasets, we introduce a smart sparse tensor generator, called \gentensor{}. This generator also enables rapid evaluation of proposed methods and algorithms without the necessity of storing the tensor.

\subsubsection{Overview}

Our generator considers the significant features of tensors. Consequently, the arpngicial tensors produced by \gentensor{} closely emulate real tensors with their respective features and characteristics. A notable advantage of this developed generator is its utilization of size-independent features, such as coefficient of variation, imbalance, and density. This allows for easy generation of instances with varying sizes, enhancing its versatility and applicability in diverse contexts.

The main idea of the proposed tensor generator is to first determine fiber counts per slice, and the number of nonzeros per fiber according to the given coefficient of variation values.
The generator has the flexibility to employ any distribution to determine these counts, yet we are currently utilizing normal or log-normal distributions to determine the nonzero layout.
The nonzero numerical values are drawn from a uniform (0,1) distribution, while The positions of the nonzeros and nonzero slice or fiber indices are selected uniformly from the corresponding index ranges.

In \gentensor{}, we provide the option to get the seed for pseudo-randomness from the user. 
Through this, one can obtain the exact same tensor when providing the same seed in different tests. 
Moreover, the user can create tensors with almost the same properties by simply changing the seed. 
The effect of seed selection will be discussed in Section \ref{sec:gen-sensitivity}.
All levels of \gentensor{} are parallelized using OpenMP for faster execution.

The generator is general and can generate any $M$-mode tensor.
Since generating random sparse tensors while simultaneously adhering to features along all modes is nearly impossible, we opt to consider the features of mode-($M-1$, $M$) slices and mode-($M-1$) fibers as inputs. 
For ease of expression, we describe the algorithm for a 3-mode tensor.

\subsubsection{The Proposed Algorithm}

The pseudocode of the proposed sparse tensor generator (\gentensor{}) is shown in Algorithm \ref{alg:tengen}. 
The generator utilizes a set of given metrics to create a tensor having these features: (i) sizes of the tensor $I_1$, $I_2$, $I_3$, (ii)  densities of slices, fibers, and nonzeros $d^{slc}$, $d^{fib}$, $d^{nz}$, (iii) coefficient of variations for fibers per slice and nonzeros per fiber $cv_{slc}^{fib}$, $cv_{fib}^{nz}$, and (iv) imbalance for fibers per slice and nonzeros per fiber $imbal_{slc}^{fib}$, $imbal_{fib}^{nz}$.

\begin{algorithm}[!t]
\begin{minipage}{0.99\linewidth}
\caption{\gentensor{}: Sparse Tensor Generator}
\label{alg:tengen}
\hspace*{\algorithmicindent} 
\hspace{-1.1em}
\textbf{Input:} $I_1$, $I_2$, $I_3$, $d^{slc}$, $d^{fib}$, $d^{nz}$, $cv_{slc}^{fib}$, $cv_{fib}^{nz}$, $imbal_{slc}^{fib}$, $imbal_{fib}^{nz}$   \\
\vspace{0.3em}
\hspace*{\algorithmicindent} 
\hspace{-1.4em}
\textbf{Output:} Tensor $\mathcal{X}$ stored in COO format

\begin{algorithmic}[1]
\vspace{0.2em}

\State Calculate $nslc$, $avg_{slc}^{fib}$, $stDev_{slc}^{fib}$,  $max_{slc}^{fib}$ using the inputs
\State $ind_{slc} \leftarrow \texttt{RandInds} (nslc, I_1 )$ \hfill $\rhd$ Array of nonzero slice indices
\vspace{0.2em}
\State $(n^{fib}_{slc}, ind_{fib} ) \leftarrow \texttt{Distribute} ( nslc, avg_{slc}^{fib}, stDev_{slc}^{fib}, max_{slc}^{fib}, I_2 )$
\hfill $\rhd$ Number and indices of fibers per slice
\State $xn^{fib}_{slc}\leftarrow \texttt{PrefixSum} ( n^{fib}_{slc} )$
\State $nfib \leftarrow xn^{fib}_{slc}(nslc) $ \hfill $\rhd$ Number of nonzero fibers
\vspace{0.2em}
\State Calculate $avg_{fib}^{nz}$, $stDev_{fib}^{nz}$,  $max_{fib}^{nz}$
\vspace{0.2em}
\State $(n^{nz}_{fib}, ind_{nz}) \leftarrow \texttt{Distribute} ( nfib,  avg_{fib}^{nz}, stDev_{fib}^{nz}, max_{fib}^{nz}, I_3 )$
\hfill $\rhd$ Number and indices of nonzeros per fiber
\State $xn^{nz}_{fib}\leftarrow \texttt{PrefixSum} ( n^{nz}_{fib} )$
\vspace{0.2em}
\For {$i \leftarrow 0$ to $nslc$}  
    \For {$j \leftarrow 0$ to $n^{fib}_{slc} ( i )$}
        \State $j' \leftarrow xn^{fib}_{slc} ( i ) + j$ \hfill $\rhd$ Global nonzero fiber index
        \For {$k \leftarrow 0$ to $n^{nz}_{fib} (j')$}
            \State $k' \leftarrow xn^{nz}_{fib} ( j' ) + k$ \hfill $\rhd$ Global nonzero index
            \State $ \mathcal{X}(k') \leftarrow \langle ind_{slc} ( i ), ind_{fib} ( i )(j), ind_{nz} ( j' )(k) \rangle$ \hfill $\rhd$ Fill nonzeros of tensor $\mathcal{X}$
        \EndFor
    \EndFor
\EndFor
\end{algorithmic}
\end{minipage}
\end{algorithm}

The algorithm first calculates the requested nonzero slice count ($nslc$), and the average, standard deviation, and maximum values for fibers per slice ($avg_{slc}^{fib}$, $stDev_{slc}^{fib}$,  $max_{slc}^{fib}$) from the given inputs.
In line 2, the nonzero slice indices are determined by the \texttt{RandInds} function, which returns $nslc$ many different indices uniformly distributed in the interval $[1,I_1]$.
In line 3, the number of fibers per slice ($n_{slc}^{fib}$) array is constructed and the indices of these nonzero fibers ($ind_{fib}$) are determined respecting the $avg_{slc}^{fib}$, $stDev_{slc}^{fib}$, and  $max_{slc}^{fib}$ values, using the function \texttt{Distribute}. 
The details of the \texttt{Distribute} method will be discussed later.
A simple prefix sum is applied on $n_{slc}^{fib}$ and we obtain $xn_{slc}^{fib}$, to be mainly used in future calculations; yet the number of nonzero fibers ($nfib$) is derived by using $xn_{slc}^{fib}$ (Lines 4-5).

We calculate the average, standard deviation, and maximum values for nonzeros per fiber ($avg_{fib}^{nz}$, $stDev_{fib}^{nz}$,  $max_{fib}^{nz}$) according to the determined $nfib$ value.
In line 7, the number of nonzeros per fiber ($n_{fib}^{nz}$) array is constructed and the indices of these nonzeros ($ind_{nz}$) are determined respecting the $avg_{fib}^{nz}$, $stDev_{fib}^{nz}$,  $max_{fib}^{nz}$ values, again using the  \texttt{Distribute} method. 
Similarly, the array $xn_{fib}^{nz}$ is obtained by applying  prefix sum on $n_{fib}^{nz}$ (Line 8).
In the last stage of the algorithm (Lines 9-17), the indices of tensor $\mathcal{X}$ are filled using the arrays $ind_{slc}$, $ind_{fib}$, and $ind_{nz}$.

The pseudocode of the \texttt{Distribute} method is given in Algorithm \ref{alg:dist}. This method takes five values as inputs, then returns a count array $cnt$ and an index array $inds$. 
We utilize the Box-Muller method \citep{golder1976box} which generates random numbers with normal distribution obeying a given standard deviation and mean.
Since this method obeys a continuous distribution, the values are real and sometimes might be negative.
However, our aim here is to construct a count array, e.g. keeping track of the nonzero count in nonzero fibers, so the target values should be positive integers. 
Thus we round the generated values to the nearest positive integers.
To avoid negative values, we switch to the log-normal distribution when needed, which guarantees positive real values.

\begin{algorithm}[!t]
\begin{minipage}{0.99\linewidth}
\caption{$\texttt{Distribute}$ }
\label{alg:dist}
\hspace*{\algorithmicindent} 
\hspace{-1.1em}
\textbf{Input:} Values $n$, $avg$, $std$, $max$, $limit$   \\
\vspace{0.3em}
\hspace*{\algorithmicindent} 
\hspace{-1.4em}
\textbf{Output:} Arrays $cnt$ and $inds$

\begin{algorithmic}[1]

\For {$i \leftarrow 0$ to $n$}
\If { $avg > 3 \times std$ }   \hfill $\rhd$ Apply normal distribution
    \State $cnt(i) \leftarrow \texttt{BoxMuller}(avg, std)$
    \vspace{0.2em}
\Else \hfill $\rhd$ Apply log-normal distribution
    \State $avg_{logNorm} \leftarrow \texttt{log}(avg^2 / (avg^2 + std^2)^{1/2})$
    \State $std_{logNorm} \leftarrow (\texttt{log}(1 + std^2 / avg^2))^{1/2}$
    \State $cnt(i) \leftarrow \texttt{exp}(\texttt{BoxMuller}(avg_{logNorm}, std_{logNorm}))$
\EndIf
\EndFor
\State $avg_{result} \leftarrow \texttt{sum}(cnt) / n$
\State $ratio = avg / avg_{result} $
\If{$ratio < 0.95$ or $ratio > 1.05$} \hfill $\rhd$ Apply scaling if needed
    \For {$i \leftarrow 0$ to $n$}
        \State $cnt(i) \leftarrow cnt(i) \times ratio $
            \EndFor
\EndIf
\For {$i \leftarrow 0$ to $n$}
    \State $cnt(i) \leftarrow \texttt{min} (cnt(i), max)$
    \State $cnt(i) \leftarrow \texttt{max} (cnt(i), 1)$
    \State $ind ( i ) \leftarrow \texttt{RandInds} (cnt( i ), limit )$ \hfill $\rhd$ Array of size $cnt(i)$, with elements in range $[1,limit]$
\EndFor
\end{algorithmic}
\end{minipage}
\end{algorithm}

The algorithm applies normal distribution if most values are expected to be positive ($avg-3\times std > 0$), or log-normal distribution, otherwise (Lines 1-8).
This choice stems from the fact that approximately 99.7\% of values sampled from a normal distribution fall within three standard deviations of the mean.
Then the algorithm scales the values of $cnt$ if the ratio of the resulting over the expected average is outside of a predetermined range, which is determined as (0.95, 1.05) in our case  (Lines 10-16).
Finally in Lines 17-21, the values of $cnt$ are adjusted to obey the minimum and maximum values, and the indices are selected uniformly using the \texttt{RandInds} method, which returns $cnt(i)$ many indices in the interval $[1,limit]$.

\subsubsection{Extension to Higher Orders}

The algorithm of \gentensor{} for higher orders is similar to the one for the 3rd-order case with the main lines.
For an $M$-mode tensor, we consider the features of mode-($M\!-\!1, M$) slices and mode-($M\!-\!1$) fibers as inputs. 
The primary additional challenge in a higher-order setting is determining the indices of nonzero slices.
It is because for an $M$-mode tensor, the nonzero slice indices themselves form a $(M-2)$-mode smaller boolean tensor, while they were simply scalars for a 3-mode tensor.
Therefore, we propose a different scheme to determine the indices of nonzero slices efficiently.

We consider four distinct cases regarding the given slice density, $d^{slc}$, which is the ratio of nonzero slice count ($nslc_{nz}$) over the total slice count ($nslc_{all}$).
If $d^{slc} > 0.97$, we round it to 1.0 and assign the slice indices in sorted order.
If $d^{slc} \in (0.5,0.97]$, we create an array 
of size $nslc_{all}$ to keep track of empty slice indices with the assumption that it can fit into memory, since $nslc_{all}$ can be at most 2 times larger than $nslc_{nz}$.
If $d^{slc} \in [0.1,0.5]$, instead of creating such an array, we traverse all possible $nslc_{all}$ indices and select a fraction of $d^{slc}$ of these indices uniformly.
Finally if $d^{slc} < 0.1$, we simply generate $nslc_{nz}$ many random 
indices for each of the respective $(M-2)$ modes.


\section{Results}
\label{sec:evaluation}

\subsection{Experimental setup}
\label{sec:exp_setup}

The experiments are conducted on an AMD EPYC 7352 CPU of 3200MHz with 512 GB of memory. It has Zen 2 microarchitecture, which includes 2 sockets, and each socket has 24 cores with 2-way simultaneous multi-threading.
Both \gentensor{} and \featensor{} are implemented in C/C++ utilizing OpenMP for shared memory parallelism and compiled with GCC using optimization level O2.

The dataset is taken from two real-world sparse tensor collections, namely FROSTT \citep{smith2017frostt} and HaTen2 \citep{jeon2015haten2}. 
We have excluded the tensors whose nonzero count is more than 1 billion.
We have also excluded the delicious-4d tensor since its fiber count exceeds the maximum value  ($1.8 \times 10^{19} $) for an unsigned long long int in the C language.
As a result, the dataset consists of 16 real-world sparse tensors whose properties are given in Table \ref{table:dataset}. Throughout the section, the runtime results are presented as an average of 3 runs.



\begin{table*}[!tbp]
\caption{Properties of the real sparse tensors from the FROSTT and HaTeN2 collections.}
\label{table:dataset}
\centering
\resizebox{0.99\textwidth}{!}{%
\begin{tabular}{lcrrrrrccccr}
\toprule
 &
   &
   &
   &
   &
   &
   &
  \multicolumn{2}{c}{Slice Count} &
  \multicolumn{2}{c}{Fiber Count} &
   \\
   \cmidrule(lr){8-9}
   \cmidrule(lr){10-11}
Name &
  \multicolumn{1}{r}{Order} &
  $I_1$ &
  $I_2$ &
  $I_3$ &
  $I_4$ &
  $I_5$ &
  All &
  Nonzero &
  All &
  Nonzero &
  NNZ \\
  \midrule
LBNL-network &
  5 &
  1,605 &
  4,198 &
  1,631 &
  4,209 &
  868,131 &
  4.1E+13 &
  1.0E+07 &
  6.9E+16 &
  6.8E+06 &
  1,698,825 \\
NIPS &
  4 &
  2,482 &
  2,862 &
  14,036 &
  17 &
  - &
  8.2E+07 &
  3.8E+06 &
  1.0E+11 &
  7.3E+06 &
  3,101,609 \\
uber &
  4 &
  183 &
  24 &
  1,140 &
  1,717 &
  - &
  2.6E+06 &
  2.6E+05 &
  4.2E+08 &
  2.8E+06 &
  3,309,490 \\
chicago-crime-comm &
  4 &
  6,186 &
  24 &
  77 &
  32 &
  - &
  8.3E+05 &
  7.1E+05 &
  3.1E+07 &
  7.8E+06 &
  5,330,673 \\
chicago-crime-geo &
  5 &
  6,185 &
  24 &
  380 &
  395 &
  32 &
  1.2E+09 &
  3.0E+07 &
  5.6E+10 &
  2.8E+07 &
  6,327,013 \\
vast-2015-mc1-3d &
  3 &
  165,427 &
  11,374 &
  2 &
  - &
  - &
  1.8E+05 &
  1.8E+05 &
  1.9E+09 &
  2.6E+07 &
  26,021,854 \\
vast-2015-mc1-5d &
  5 &
  165,427 &
  11,374 &
  2 &
  100 &
  89 &
  3.6E+11 &
  1.2E+08 &
  1.7E+13 &
  1.0E+08 &
  26,021,945 \\
DARPA1998 &
  3 &
  22,476 &
  22,476 &
  23,776,223 &
  - &
  - &
  2.4E+07 &
  2.4E+07 &
  1.1E+12 &
  5.5E+07 &
  28,436,033 \\
enron &
  4 &
  6,066 &
  5,699 &
  244,268 &
  1,176 &
  - &
  3.2E+09 &
  2.2E+07 &
  1.2E+13 &
  9.0E+07 &
  54,202,099 \\
NELL-2 &
  3 &
  12,092 &
  9,184 &
  28,818 &
  - &
  - &
  5.0E+04 &
  5.0E+04 &
  7.2E+08 &
  3.8E+07 &
  76,879,419 \\
freebase\_music &
  3 &
  23,343,790 &
  23,344,784 &
  166 &
  - &
  - &
  4.7E+07 &
  4.6E+07 &
  5.4E+14 &
  2.2E+08 &
  99,546,551 \\
flickr-3d &
  3 &
  319,686 &
  28,153,045 &
  1,607,191 &
  - &
  - &
  3.0E+07 &
  3.0E+07 &
  5.5E+13 &
  1.5E+08 &
  112,890,310 \\
flickr-4d &
  4 &
  319,686 &
  28,153,045 &
  1,607,191 &
  731 &
  - &
  5.5E+13 &
  1.9E+08 &
  1.5E+19 &
  2.8E+08 &
  112,890,310 \\
freebase\_sampled & 3 & 38,954,435 & 38,955,429 & 532       & -     & - & 7.8E+07 & 7.4E+07 & 1.5E+15 & 3.2E+08 & 139,920,771 \\
delicious-3d      & 3 & 532,924    & 17,262,471 & 2,480,308 & -     & - & 2.0E+07 & 2.0E+07 & 5.3E+13 & 1.6E+08 & 140,126,181 \\
NELL-1 &
  3 &
  2,902,330 &
  2,143,368 &
  25,495,389 &
  - &
  - &
  3.1E+07 &
  3.1E+07 &
  1.3E+14 &
  2.5E+08 &
  143,599,552 \\
\bottomrule
\end{tabular}%
}
\end{table*}

\subsection{Performance of Feature Extraction Methods}
\label{sec:feat-reasult}

The performance of four different feature extraction methods in \featensor{} is compared. 
For this experiment, we use the maximum number of available hardware threads in the machine, which is 96.
Since our dataset includes both 3rd-order and higher-order tensors, for a fair comparison, we run all four methods in \featensor{} with the only-3-mode option, i.e. we extract the features corresponding to modes with the largest three sizes.
The time measurement covers extracting all features along all modes, including the preprocessing time for each method, e.g. sorting time for the sort-based method and time for preparing the hash table for the hash-based method.
The experiments validate that all feature extraction methods in $\featensor{}$ give the same and exact results as feature sets, therefore we do not present an accuracy comparison across different methods.

\begin{figure*}[!tbp]
    \begin{center}
    \includegraphics[width=0.99\linewidth]{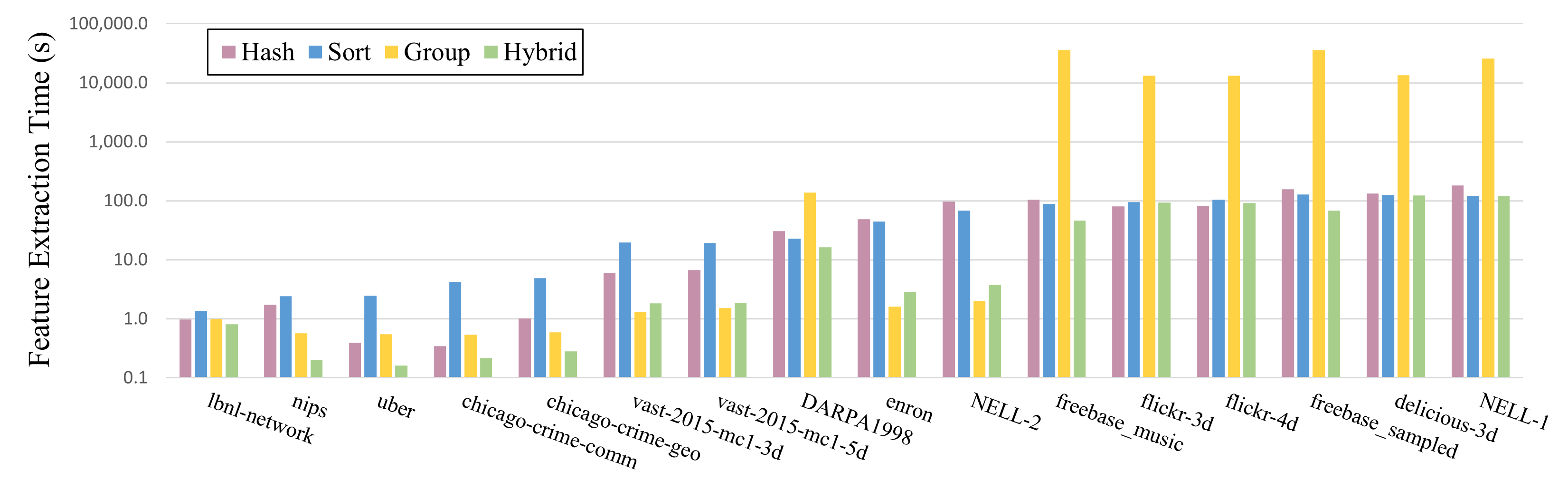}
    \caption{Runtime comparison for different feature extraction methods.}
    \label{fig:feat-result}
    \end{center}
\end{figure*}

Table \ref{fig:feat-result} presents a detailed runtime comparison between the sort-based and group-based methods at the mode level.
The best runtime value for each mode is shown in bold. 
We also report the value of the decision metric ($size(m)\times size(m+1)$) in the hybrid approach for each mode $m$. 
The tensors are shown in increasing order of nonzero counts.
It is seen that the superiority of a method varies between different modes within the execution of each tensor. Although the grouping-based method tends to be costly in total time due to some modes with larger sizes, it is still the winner for some modes of even large tensors. This is the main reason why integrating the grouping-based method for specific modes within the hybrid approach yields improved performance.
We observe that the grouping method yields the best performance when the corresponding decision metric for that mode is less than $10^{11}$ for all cases except LBNL-network, which is the tensor with the smallest nonzero count in our dataset.
Therefore, we empirically set the threshold value as $\lambda=10^{11}$ in the hybrid method. This strategy prevents excessive cost in high-dimensional modes and enables faster execution by combining the strengths of both methods.



\begin{table}[!tbp] 
\caption{
Comparison of sort- and group-based methods in terms of mode-level runtime in seconds, where best results for each mode are shown in bold; along with decision metric ($size(m)\times size(m+1)$) values of hybrid method for each mode $m$, where values less than $\lambda=10^{11}$ are shown in bold.}
\label{table:sort-group-time}
\centering
\resizebox{0.85\columnwidth}{!}{%
\begin{tabular}{lrrrrrrrrr}
\toprule
 & \multicolumn{3}{c}{Sort-based} & \multicolumn{3}{c}{Group-based} & \multicolumn{3}{c}{Decision metric in hybrid method} \\
 \cmidrule(lr){2-4}
                \cmidrule(lr){5-7}
                 \cmidrule(lr){8-10}
Tensor & mode-1 & mode-2 & mode-3 & mode-1 & mode-2 & mode-3 & mode-1 & mode-2 & mode-3 \\
\midrule
LBNL-network & 0.60 & {\color[HTML]{CC00FF} \textbf{0.37}} & {\color[HTML]{CC00FF} \textbf{0.35}} & {\color[HTML]{00B050} \textbf{0.32}} & 0.60 & 0.61 & {\color[HTML]{00B0F0} \textbf{1.8E+07}} & {\color[HTML]{00B0F0} \textbf{3.6E+09}} & {\color[HTML]{00B0F0} \textbf{3.7E+09}} \\
NIPS & 1.03 & 0.69 & 0.65 & {\color[HTML]{00B050} \textbf{0.11}} & {\color[HTML]{00B050} \textbf{0.14}} & {\color[HTML]{00B050} \textbf{0.15}} & {\color[HTML]{00B0F0} \textbf{7.1E+06}} & {\color[HTML]{00B0F0} \textbf{3.5E+07}} & {\color[HTML]{00B0F0} \textbf{4.0E+07}} \\
uber & 1.07 & 0.68 & 0.67 & {\color[HTML]{00B050} \textbf{0.09}} & {\color[HTML]{00B050} \textbf{0.10}} & {\color[HTML]{00B050} \textbf{0.12}} & {\color[HTML]{00B0F0} \textbf{2.1E+05}} & {\color[HTML]{00B0F0} \textbf{3.1E+05}} & {\color[HTML]{00B0F0} \textbf{2.0E+06}} \\
chicago-crime-comm & 1.75 & 1.20 & 1.16 & {\color[HTML]{00B050} \textbf{0.11}} & {\color[HTML]{00B050} \textbf{0.13}} & {\color[HTML]{00B050} \textbf{0.14}} & {\color[HTML]{00B0F0} \textbf{2.5E+03}} & {\color[HTML]{00B0F0} \textbf{2.0E+05}} & {\color[HTML]{00B0F0} \textbf{4.8E+05}} \\
chicago-crime-geo & 2.09 & 1.34 & 1.30 & {\color[HTML]{00B050} \textbf{0.17}} & {\color[HTML]{00B050} \textbf{0.18}} & {\color[HTML]{00B050} \textbf{0.19}} & {\color[HTML]{00B0F0} \textbf{1.5E+05}} & {\color[HTML]{00B0F0} \textbf{2.4E+06}} & {\color[HTML]{00B0F0} \textbf{2.4E+06}} \\
vast-2015-mc1-3d & 8.26 & 5.71 & 5.18 & {\color[HTML]{00B050} \textbf{0.60}} & {\color[HTML]{00B050} \textbf{0.82}} & {\color[HTML]{00B050} \textbf{0.85}} & {\color[HTML]{00B0F0} \textbf{2.3E+04}} & {\color[HTML]{00B0F0} \textbf{3.3E+05}} & {\color[HTML]{00B0F0} \textbf{1.9E+09}} \\
vast-2015-mc1-5d & 8.00 & 5.39 & 5.39 & {\color[HTML]{00B050} \textbf{0.78}} & {\color[HTML]{00B050} \textbf{0.85}} & {\color[HTML]{00B050} \textbf{1.01}} & {\color[HTML]{00B0F0} \textbf{1.1E+06}} & {\color[HTML]{00B0F0} \textbf{1.7E+07}} & {\color[HTML]{00B0F0} \textbf{1.9E+09}} \\
DARPA1998 & 8.91 & {\color[HTML]{CC00FF} \textbf{6.16}} & {\color[HTML]{CC00FF} \textbf{6.73}} & {\color[HTML]{00B050} \textbf{1.30}} & 103.10 & 137.98 & {\color[HTML]{00B0F0} \textbf{5.1E+08}} & 5.3E+11 & 5.3E+11 \\
enron & 16.70 & 14.58 & 12.37 & {\color[HTML]{00B050} \textbf{0.68}} & {\color[HTML]{00B050} \textbf{0.79}} & {\color[HTML]{00B050} \textbf{1.19}} & {\color[HTML]{00B0F0} \textbf{3.5E+07}} & {\color[HTML]{00B0F0} \textbf{1.4E+09}} & {\color[HTML]{00B0F0} \textbf{1.5E+09}} \\
NELL-2 & 26.04 & 20.83 & 19.07 & {\color[HTML]{00B050} \textbf{0.84}} & {\color[HTML]{00B050} \textbf{0.85}} & {\color[HTML]{00B050} \textbf{1.51}} & {\color[HTML]{00B0F0} \textbf{1.1E+08}} & {\color[HTML]{00B0F0} \textbf{2.6E+08}} & {\color[HTML]{00B0F0} \textbf{3.5E+08}} \\
freebase\_music & 34.42 & 28.64 & {\color[HTML]{CC00FF} \textbf{21.99}} & {\color[HTML]{00B050} \textbf{3.44}} & {\color[HTML]{00B050} \textbf{16.72}} & 36,000.00 & {\color[HTML]{00B0F0} \textbf{3.9E+09}} & {\color[HTML]{00B0F0} \textbf{3.9E+09}} & 5.4E+14 \\
flickr-3d & {\color[HTML]{CC00FF} \textbf{38.23}} & {\color[HTML]{CC00FF} \textbf{29.81}} & {\color[HTML]{CC00FF} \textbf{24.63}} & 318.65 & 4,846.71 & 13,158.43 & 5.1E+11 & 9.0E+12 & 4.5E+13 \\
flickr-4d & {\color[HTML]{CC00FF} \textbf{38.87}} & {\color[HTML]{CC00FF} \textbf{39.14}} & {\color[HTML]{CC00FF} \textbf{24.39}} & 326.67 & 4,798.01 & 13,109.18 & 5.1E+11 & 9.0E+12 & 4.5E+13 \\
freebase\_sampled & 50.63 & 42.63 & {\color[HTML]{CC00FF} \textbf{32.03}} & {\color[HTML]{00B050} \textbf{8.31}} & {\color[HTML]{00B050} \textbf{28.43}} & 36,000.00 & {\color[HTML]{00B0F0} \textbf{2.1E+10}} & {\color[HTML]{00B0F0} \textbf{2.1E+10}} & 1.5E+15 \\
delicious-3d & {\color[HTML]{CC00FF} \textbf{51.05}} & {\color[HTML]{CC00FF} \textbf{39.65}} & {\color[HTML]{CC00FF} \textbf{31.43}} & 1,209.29 & 4,840.33 & 13,517.24 & 1.3E+12 & 9.2E+12 & 4.3E+13 \\
NELL-1 & {\color[HTML]{CC00FF} \textbf{47.36}} & {\color[HTML]{CC00FF} \textbf{38.25}} & {\color[HTML]{CC00FF} \textbf{33.05}} & 4,346.74 & 21,478.60 & 25,525.99 & 6.2E+12 & 5.5E+13 & 7.4E+13\\
\bottomrule
\end{tabular}%
}
\end{table}

Figure \ref{fig:feat-result} shows the total feature extraction time for each method and each tensor. 
Excluding the hybrid method,
the grouping-based method is superior for smaller tensors, whereas the sort-based method is better for larger tensors.
Overall, the hybrid method is the best-performing one for most of the tensors, and it still ranks as the second-best method for the remaining tensors.
The results confirm that the hybrid method balances sorting and grouping methods by selecting the appropriate strategy at a mode level according to mode-wise characteristics.



\subsection{Performance of Tensor Generator}

We evaluate the effectiveness of the proposed tensor generator in terms of feature quality, tensor decomposition performance, and sensitivity to seed selection.
Except for the sensitivity analysis part, we use the generated tensors in which the seed for pseudo-randomness is set to 0.
The generated tensors are created by \gentensor{} using the features of real tensors, which are obtained via our feature extractor \featensor{}.
Note that \gentensor{} needs the fiber and slice features from only a single mode, irrespective of the order of the tensor. 
Since all the methods in \featensor{} give the same features exactly but only their runtime differ, any of the methods in \featensor{} can be used to obtain real tensor features as inputs for \gentensor{}.

\subsubsection{Feature Quality}
\label{sec:gen-feat}

Table \ref{table:generator} shows the comparison of the generated tensors with their original versions, i.e. real tensors, in terms of some important features. 
We present the features of the original and the respective generated tensors, as well as the ratio of the resulting value of the generated tensor over that of the original tensor.  
The ratio values are colored green if the value is between 0.9 and 1.1; red if it is less than 0.5 or more than 2; and orange, otherwise.
For zero or too small (less than 0.1) coefficient of variation values, the generator often yields values with zero coefficient of variation.
For those cases, the ratio values are omitted from the table since the ratio will appear as either zero or undefined (0/0).

As can be seen in the table, the resulting densities, i.e. the nonzero count of the generated tensors, are at least 0.96 times smaller or at most 1.05 times larger than the ones of the respective original tensors.
The success of the generator in terms of obeying the given density is seen in both levels of nonzero slice, nonzero fiber, and nonzero density.

In methods that generate values to obey a given density and variation, there is a trade-off between strictly obeying the density or the variation. 
In other words, if one prioritizes achieving the exact coefficient of variation, the density might get far from the desired value.
However, since the number of nonzeros is the most significant feature that a generator must obey for performance concerns, we opted to prioritize adhering to density.
For this, in \gentensor{} we apply some scalings during calculations to catch the given density.
It is the reason why the resulting ratios in coefficient of variation seem relatively low compared to the ratios in densities.

We also present the tensor generation times in the last column of Table \ref{table:generator}. These are the runtimes of \gentensor{} in seconds when working with the maximum available thread count, which is 96 in our case. 
We observe that \gentensor{} takes only a few seconds when generating medium-size tensors.
To be precise, it takes less than 5 seconds for 9 out of 16 tensors.
Only for some large tensors (4 cases in our dataset), the runtime of \gentensor{} can go up to a few hours. 
Note that the execution of \gentensor{} depends on the requested nonzero count as well as the slice and fiber counts; thus the sizes of the tensor also affect the execution time.



\begin{table*}[!tbp]
\caption{
Feature comparison of the generated tensors with their original versions; and time to generate tensors.}
\label{table:generator}
\centering
\resizebox{\textwidth}{!}{%
\begin{tabular}{lrrcrrcrrcccccccr}
\bottomrule
 &
  \multicolumn{6}{c}{Coefficient of   Variation} &
  \multicolumn{9}{c}{Density} \\
\cmidrule(lr){2-7}
\cmidrule(lr){8-16}
 &
  \multicolumn{3}{c}{Fiber per Slice} &
  \multicolumn{3}{c}{Nonzero per Fiber} &
  \multicolumn{3}{c}{Nonzero Slice} &
  \multicolumn{3}{c}{Nonzero Fiber} &
  \multicolumn{3}{c}{Nonzero} \\
   \cmidrule(lr){2-4}
    \cmidrule(lr){5-7}
     \cmidrule(lr){8-10}
      \cmidrule(lr){11-13}
       \cmidrule(lr){14-16}
\multirow{-3}{*}{Name} &
  Org &
  Gen &
  Ratio &
  Org &
  Gen &
  Ratio &
  Org &
  Gen &
  Ratio &
  Org &
  Gen &
  Ratio &
  Org &
  Gen &
  Ratio &
  \multirow{-3}{*}{Time (s)}
  \\
  \midrule
LBNL-network &
  8.0 &
  4.5 &
  {\color[HTML]{BF8F00} \textbf{0.56}} &
  26.2 &
  12.1 &
  {\color[HTML]{C65911} \textbf{0.46}} &
  2.1E-06 &
  2.1E-06 &
  {\color[HTML]{548235} \textbf{1.00}} &
  8.4E-10 &
  9.5E-10 &
  {\color[HTML]{BF8F00} \textbf{1.13}} &
  4.2E-14 &
  4.4E-14 &
  {\color[HTML]{548235} \textbf{1.05}} &
  1.0 \\
NIPS &
  0.2 &
  0.2 &
  {\color[HTML]{548235} \textbf{0.98}} &
  0.0 &
  0.0 &
  {\color[HTML]{548235} \textbf{-}} &
  8.3E-04 &
  8.3E-04 &
  {\color[HTML]{548235} \textbf{1.00}} &
  3.1E-05 &
  3.0E-05 &
  {\color[HTML]{548235} \textbf{0.98}} &
  1.8E-06 &
  1.8E-06 &
  {\color[HTML]{548235} \textbf{0.98}} &
  0.3 \\
uber &
  0.2 &
  0.2 &
  {\color[HTML]{548235} \textbf{0.94}} &
  1.0 &
  1.0 &
  {\color[HTML]{548235} \textbf{0.94}} &
  1.0E+00 &
  1.0E+00 &
  {\color[HTML]{548235} \textbf{1.00}} &
  1.4E-01 &
  1.3E-01 &
  {\color[HTML]{548235} \textbf{0.93}} &
  3.8E-04 &
  3.8E-04 &
  {\color[HTML]{548235} \textbf{0.99}} &
  0.1 \\
chicago-crime-comm &
  0.4 &
  0.3 &
  {\color[HTML]{BF8F00} \textbf{0.82}} &
  0.5 &
  0.4 &
  {\color[HTML]{BF8F00} \textbf{0.82}} &
  9.5E-01 &
  9.5E-01 &
  {\color[HTML]{548235} \textbf{1.00}} &
  3.2E-01 &
  2.7E-01 &
  {\color[HTML]{BF8F00} \textbf{0.83}} &
  1.5E-02 &
  1.4E-02 &
  {\color[HTML]{548235} \textbf{0.96}} &
  0.3 \\
chicago-crime-geo &
  0.3 &
  0.3 &
  {\color[HTML]{548235} \textbf{0.94}} &
  0.1 &
  0.0 &
  {\color[HTML]{C65911} \textbf{-}} &
  1.0E-01 &
  1.0E-01 &
  {\color[HTML]{548235} \textbf{1.00}} &
  2.8E-04 &
  2.9E-04 &
  {\color[HTML]{548235} \textbf{1.01}} &
  8.9E-06 &
  8.9E-06 &
  {\color[HTML]{548235} \textbf{1.01}} &
  1.6 \\
vast-2015-mc1-3d &
  0.5 &
  0.5 &
  {\color[HTML]{548235} \textbf{1.00}} &
  0.0 &
  0.0 &
  {\color[HTML]{548235} \textbf{-}} &
  1.0E+00 &
  1.0E+00 &
  {\color[HTML]{548235} \textbf{1.00}} &
  1.4E-02 &
  1.4E-02 &
  {\color[HTML]{548235} \textbf{0.99}} &
  6.9E-03 &
  6.9E-03 &
  {\color[HTML]{548235} \textbf{0.99}} &
  2.2 \\
vast-2015-mc1-5d &
  0.0 &
  0.0 &
  {\color[HTML]{C65911} \textbf{-}} &
  0.0 &
  0.0 &
  {\color[HTML]{C65911} \textbf{-}} &
  6.9E-03 &
  6.9E-03 &
  {\color[HTML]{548235} \textbf{1.00}} &
  6.9E-05 &
  6.9E-05 &
  {\color[HTML]{548235} \textbf{1.00}} &
  7.8E-07 &
  7.8E-07 &
  {\color[HTML]{548235} \textbf{1.00}} &
  4.8 \\
DARPA1998 &
  13.1 &
  8.2 &
  {\color[HTML]{BF8F00} \textbf{0.63}} &
  23.1 &
  14.0 &
  {\color[HTML]{BF8F00} \textbf{0.61}} &
  8.0E-01 &
  8.1E-01 &
  {\color[HTML]{548235} \textbf{1.02}} &
  1.5E-04 &
  1.6E-04 &
  {\color[HTML]{548235} \textbf{1.03}} &
  2.4E-09 &
  2.4E-09 &
  {\color[HTML]{548235} \textbf{1.00}} &
  147.8 \\
enron &
  4.1 &
  3.6 &
  {\color[HTML]{BF8F00} \textbf{0.87}} &
  1.8 &
  1.4 &
  {\color[HTML]{BF8F00} \textbf{0.76}} &
  4.4E-03 &
  4.4E-03 &
  {\color[HTML]{548235} \textbf{1.00}} &
  3.7E-06 &
  3.7E-06 &
  {\color[HTML]{548235} \textbf{0.99}} &
  5.5E-09 &
  5.7E-09 &
  {\color[HTML]{548235} \textbf{1.05}} &
  3.1 \\
NELL-2 &
  3.3 &
  3.1 &
  {\color[HTML]{548235} \textbf{0.94}} &
  0.9 &
  1.1 &
  {\color[HTML]{BF8F00} \textbf{1.25}} &
  1.0E+00 &
  1.0E+00 &
  {\color[HTML]{548235} \textbf{1.00}} &
  3.0E-03 &
  3.1E-03 &
  {\color[HTML]{548235} \textbf{1.01}} &
  2.4E-05 &
  2.4E-05 &
  {\color[HTML]{548235} \textbf{0.99}} &
  0.6 \\
freebase\_music &
  24.4 &
  20.2 &
  {\color[HTML]{BF8F00} \textbf{0.83}} &
  0.1 &
  0.0 &
  {\color[HTML]{C65911} \textbf{-}} &
  9.7E-01 &
  1.0E+00 &
  {\color[HTML]{548235} \textbf{1.03}} &
  1.8E-07 &
  1.9E-07 &
  {\color[HTML]{548235} \textbf{1.04}} &
  1.1E-09 &
  1.1E-09 &
  {\color[HTML]{548235} \textbf{1.03}} &
  4,370.6 \\
flickr-3d &
  3.3 &
  3.2 &
  {\color[HTML]{548235} \textbf{0.97}} &
  1.0 &
  1.0 &
  {\color[HTML]{548235} \textbf{0.99}} &
  1.0E+00 &
  1.0E+00 &
  {\color[HTML]{548235} \textbf{1.00}} &
  3.1E-06 &
  3.1E-06 &
  {\color[HTML]{548235} \textbf{1.00}} &
  7.8E-12 &
  7.9E-12 &
  {\color[HTML]{548235} \textbf{1.01}} &
  2,080.4 \\
flickr-4d &
  1.0 &
  1.0 &
  {\color[HTML]{548235} \textbf{0.99}} &
  0.0 &
  0.0 &
  {\color[HTML]{548235} \textbf{-}} &
  3.1E-06 &
  3.1E-06 &
  {\color[HTML]{548235} \textbf{1.00}} &
  7.8E-12 &
  7.9E-12 &
  {\color[HTML]{548235} \textbf{1.01}} &
  1.1E-14 &
  1.1E-14 &
  {\color[HTML]{548235} \textbf{1.01}} &
  1,180.8 \\
freebase\_sampled &
  24.0 &
  19.4 &
  {\color[HTML]{BF8F00} \textbf{0.81}} &
  0.1 &
  0.0 &
  {\color[HTML]{C65911} \textbf{-}} &
  9.1E-01 &
  9.1E-01 &
  {\color[HTML]{548235} \textbf{1.00}} &
  9.2E-08 &
  9.6E-08 &
  {\color[HTML]{548235} \textbf{1.05}} &
  1.7E-10 &
  1.8E-10 &
  {\color[HTML]{548235} \textbf{1.04}} &
  10,608.6 \\
delicious-3d &
  2.8 &
  2.7 &
  {\color[HTML]{548235} \textbf{0.99}} &
  1.4 &
  1.0 &
  {\color[HTML]{BF8F00} \textbf{0.71}} &
  1.0E+00 &
  1.0E+00 &
  {\color[HTML]{548235} \textbf{1.00}} &
  5.1E-06 &
  5.1E-06 &
  {\color[HTML]{548235} \textbf{1.00}} &
  6.1E-12 &
  6.1E-12 &
  {\color[HTML]{548235} \textbf{1.00}} &
  6,332.0 \\
NELL-1 &
  13.6 &
  10.8 &
  {\color[HTML]{BF8F00} \textbf{0.80}} &
  7.5 &
  4.5 &
  {\color[HTML]{BF8F00} \textbf{0.60}} &
  1.0E+00 &
  1.0E+00 &
  {\color[HTML]{548235} \textbf{1.00}} &
  2.8E-06 &
  2.8E-06 &
  {\color[HTML]{548235} \textbf{1.01}} &
  9.1E-13 &
  9.2E-13 &
  {\color[HTML]{548235} \textbf{1.01}} &
  23,384.5 \\
  \bottomrule
\end{tabular}%
}
\end{table*}

\subsubsection{CPD Performance}

We evaluate the effectiveness of the generated tensors in mimicking the behavior of real tensors, particularly in terms of tensor decomposition performance.
For this, the performance of the generated tensors is compared with the performance of the naive random tensors, which have the same sizes and nonzero counts 
as the original tensors but the nonzero locations are uniformly random.
The SPLATT \citep{smith2015splatt} and ParTI \citep{li2018parti} tools are used for applying the CPD decomposition. 
Both tools are constrained to take the same (50) number of iterations for a fair comparison and the time of the first 5 iterations are ignored for cache warp-up.
For each case, we take 5 independent runs and choose the minimum runtime to represent the peak performance of the system and to be less susceptible to noise than the average.

Figure \ref{fig:cpd} illustrates the comparison between the generated and naive random tensors in their ability to resemble real tensors regarding the CPD performance.
The runtime results for both the naive random and the generated tensors are normalized with respect to the runtime obtained for the respective original tensor.
Therefore, the normalized values closer to 1.0 are interpreted as more successful in terms of resembling the original tensor performance. 



\setcounter{figure}{4}
\setcounter{subfigure}{0}
\begin{subfigure}
\setcounter{figure}{4}
\setcounter{subfigure}{0}
    \centering
    \begin{minipage}[b]{0.95\textwidth}
        \includegraphics[width=\linewidth]{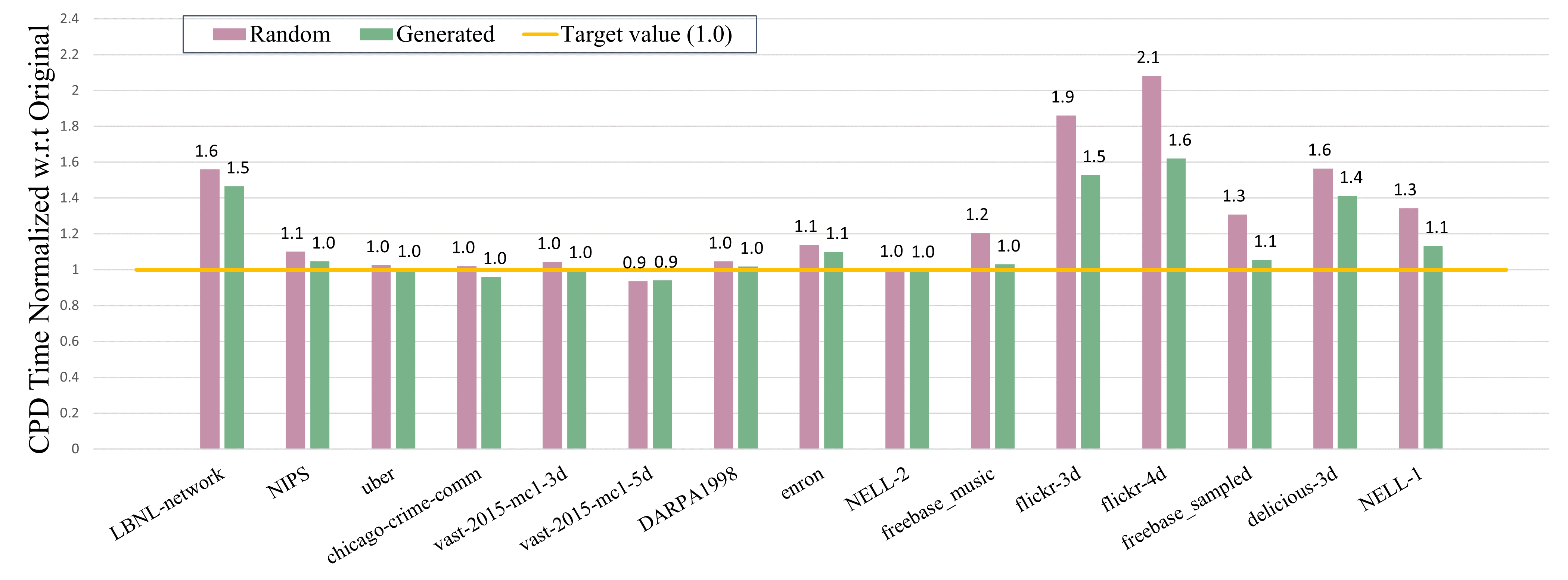}
        \caption{ParTI performance comparison for a single thread}
        \label{fig:parti-1}
    \end{minipage}  
   
\setcounter{figure}{4}
\setcounter{subfigure}{1}
    \begin{minipage}[b]{0.95\textwidth}
        \includegraphics[width=\linewidth]{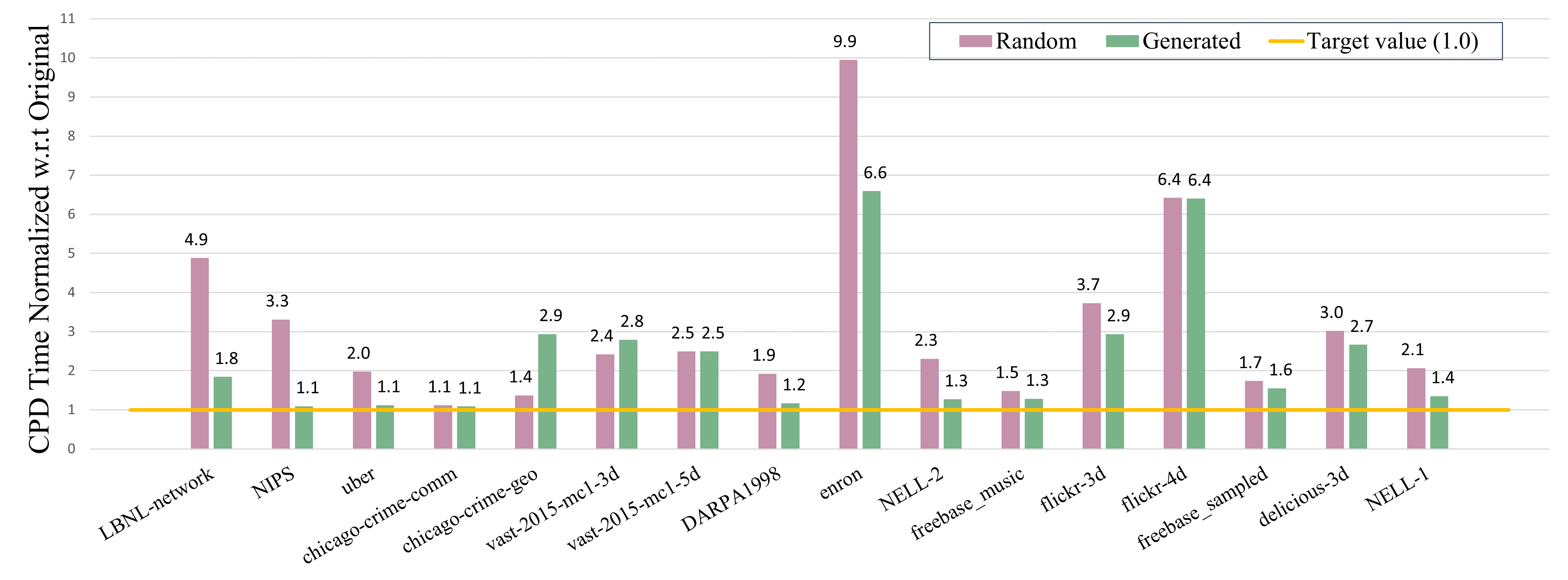}
        \caption{SPLATT performance comparison for a single thread}
        \label{fig:splatt-1}
    \end{minipage}

\setcounter{figure}{4}
\setcounter{subfigure}{2}
    \begin{minipage}[b]{0.95\textwidth}
        \includegraphics[width=\linewidth]{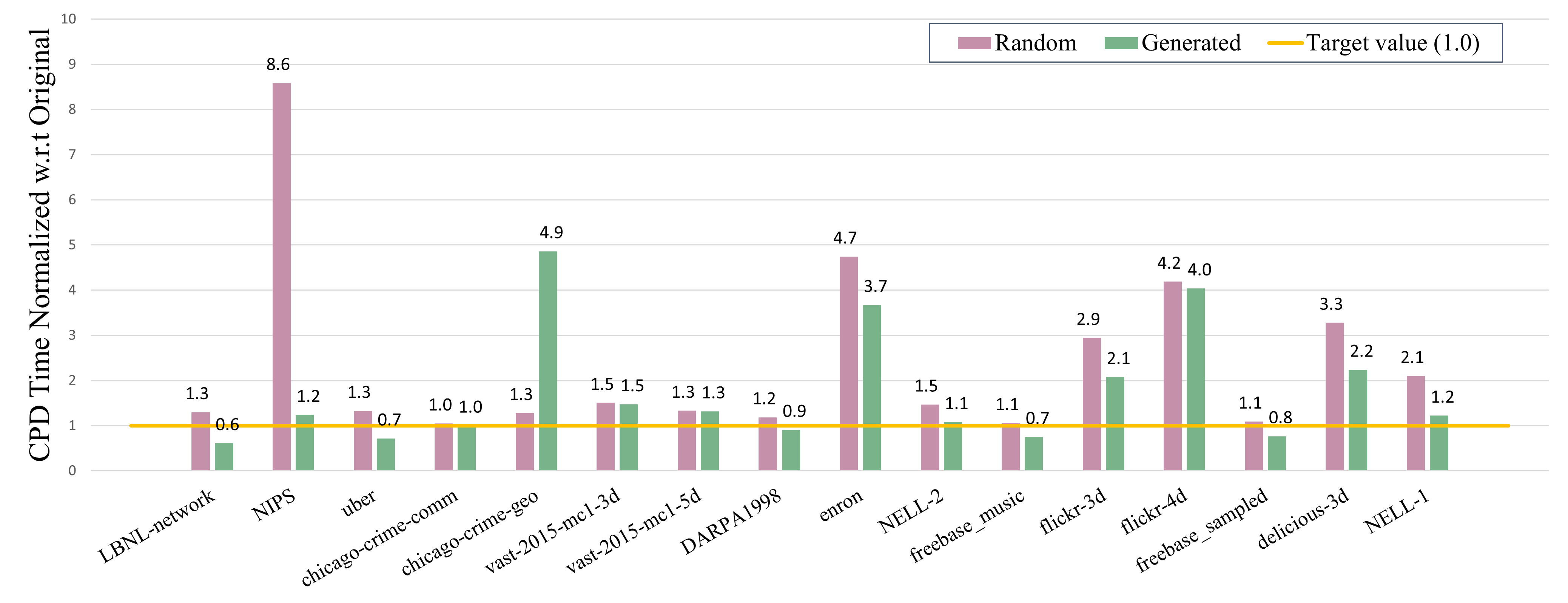}
        \caption{SPLATT performance comparison for 48 threads}
        \label{fig:splatt-48}
    \end{minipage}

\setcounter{figure}{4}
\setcounter{subfigure}{-1}
    \caption{CPD performance comparison of the naive random and the generated tensors. The values are normalized with respect to the runtime obtained for the respective original tensor.}
    \label{fig:cpd}
\end{subfigure}

Figure \ref{fig:parti-1} depicts the CPD performance comparison by using the ParTI tool with a single thread.
It is evident that the generated tensors emerge as the clear winner in most cases, and in the remaining scenarios, they are comparable to the naive random tensors but never inferior. 
The superiority of the generated tensors over the naive random ones is especially higher for larger tensors.
The ParTI tool gave an error for the chicago-crime-geo tensor, so this tensor is not presented in Figure \ref{fig:parti-1} and the rest of the experiments are conducted by using only the SPLATT tool.

In Figure \ref{fig:splatt-1}, we present the CPD performance comparison by using the SPLATT tool with a single thread.
As seen in the figure, the generated tensors show significantly better performance than the naive random ones for 11 out of 16 cases; and yield similar performance with naive random tensors for 3 tensors in the serial setting.
For instance, for the NIPS tensor, while the naive random tensor is 3.3 times slower than the original tensor, the tensor generated with \gentensor{} yields a CPD time of only 1.1 times more than the runtime of the original real tensor.

Figure \ref{fig:splatt-48} shows the CPD performance comparison by using SPLATT with 48 threads.
We observe that the generated tensors are superior to the naive random ones for 8 cases; and inferior for only 1 tensor (chicago-crime-geo). 
The generated tensors show almost the same performance as the naive random ones for 3 cases but for those, their performance is either already the same as the respective real tensor (chicago-crime-comm), or only 1.3 (vast-2015-mc1-5d) and 1.5 (vast-2015-mc1-3d) times far from the performance of the original tensors. 

Although these results demonstrate that \gentensor{} performs well in most cases, there are a few instances where the generated tensors differ from real tensors in performance. This discrepancy can be attributed to differences in tensor values. Since CPD performance is influenced not only by the sparsity pattern but also by the rank of the tensors, which is an aspect beyond the scope of this work, these value-related differences explain the observed variations.
For example, in the case of the chicago-crime-geo tensor, the original nonzeros represent count data and follow a highly skewed distribution. On the other hand, \gentensor{} assigns nonzeros from a uniform (0,1) distribution. Although structurally consistent, this difference in value distribution may result in different numerical scaling and optimization behavior, which explains the deviation in CPD performance observed with SPLATT. This illustrates how even structurally similar tensors may diverge in runtime depending on their value statistics.

\subsubsection{Sensitivity Analysis}
\label{sec:gen-sensitivity}

To evaluate the sensitivity of our generator, we produce 3 different versions for each generated tensor by taking the seed for pseudo-randomness as 0, 1, and 2.
Table \ref{table:sensitivity} shows the nonzero counts and the CPD (SPLATT) runtime for 1 and 48 threads (in seconds).
We present the coefficient of variation (CV) for each case. 
As can be seen in the table, the CV for nonzero count is at most $5 \times 10^{-3}$, whereas the CV for CPD performance is at most 0.1.
These results demonstrate that the generator is stable across different random seeds and produces consistent sparsity patterns and decomposition performance.
This stability makes \gentensor{} reliable for repeatable experimentation and benchmarking scenarios. 



\begin{table*}[!t]
\centering
\caption{
Comparison of the generated tensors with different seeds in terms of nonzero count and CPD time in seconds. }
\label{table:sensitivity}
\resizebox{\textwidth}{!}{%
\begin{tabular}{lrrrrrrrrrrrr}
\toprule
\multicolumn{1}{c}{} & \multicolumn{4}{c}{Nonzero Count}                 & \multicolumn{4}{c}{CPD Time (1 Thread)} & \multicolumn{4}{c}{CPD Time (48 Threads)} \\
\cmidrule(lr){2-5}
\cmidrule(lr){6-9}
\cmidrule(lr){10-13}
Name               & Seed=0      & Seed=1      & Seed=2      & CV      & Seed=0  & Seed=1  & Seed=2  & CV      & Seed=0 & Seed=1 & Seed=2 & CV      \\
\midrule
LBNL-network       & 1,699,088   & 1,699,162   & 1,699,444   & 9.0E-05 & 21.3    & 21.3    & 22.1    & 1.8E-02 & 3.9    & 5.0    & 4.9    & 1.1E-01 \\
NIPS               & 3,041,941   & 3,042,054   & 3,041,816   & 3.2E-05 & 6.1     & 6.1     & 6.7     & 3.9E-02 & 0.5    & 0.5    & 0.6    & 4.2E-02 \\
uber               & 3,283,106   & 3,282,919   & 3,283,336   & 5.2E-05 & 6.4     & 6.3     & 6.3     & 6.1E-03 & 0.3    & 0.3    & 0.3    & 0.0E+00 \\
chicago-crime-comm & 5,131,130   & 5,131,402   & 5,131,234   & 2.2E-05 & 8.0     & 7.7     & 7.9     & 1.5E-02 & 0.4    & 0.4    & 0.4    & 7.0E-03 \\
chicago-crime-geo  & 6,364,282   & 6,364,306   & 6,365,068   & 5.7E-05 & 75.3    & 100.8   & 90.9    & 1.2E-01 & 7.9    & 9.8    & 8.8    & 8.9E-02 \\
vast-2015-mc1-3d   & 25,800,284  & 25,802,581  & 25,801,506  & 3.6E-05 & 64.1    & 67.1    & 70.4    & 3.8E-02 & 29.1   & 31.3   & 29.7   & 3.1E-02 \\
vast-2015-mc1-5d   & 26,021,841  & 26,021,841  & 26,021,841  & 0.0E+00 & 268.9   & 269.6   & 268.9   & 1.2E-03 & 218.4  & 214.2  & 212.2  & 1.2E-02 \\
DARPA1998          & 28,732,452  & 28,406,813  & 28,405,526  & 5.4E-03 & 250.7   & 251.5   & 269.8   & 3.4E-02 & 84.6   & 94.6   & 93.0   & 4.8E-02 \\
enron              & 56,967,347  & 57,055,466  & 57,032,145  & 6.5E-04 & 588.9   & 670.4   & 651.7   & 5.5E-02 & 29.8   & 32.7   & 32.7   & 4.3E-02 \\
NELL-2             & 76,203,551  & 76,202,209  & 76,203,884  & 9.5E-06 & 89.8    & 89.5    & 90.1    & 2.6E-03 & 7.7    & 7.9    & 7.8    & 1.4E-02 \\
freebase\_music      & 102,877,680 & 102,916,007 & 102,874,661 & 1.8E-04 & 1,459.2  & 1,495.5  & 1,480.7 & 1.0E-02 & 243.3    & 316.4    & 312.1   & 1.2E-01   \\
flickr-3d          & 113,727,282 & 113,736,842 & 113,674,701 & 2.4E-04 & 1,471.7 & 1,458.7 & 1,544.7 & 2.5E-02 & 261.1  & 292.3  & 308.5  & 6.8E-02 \\
flickr-4d          & 113,641,027 & 113,641,445 & 113,640,707 & 2.7E-06 & 4,714.5 & 4,786.8 & 4,700.2 & 8.0E-03 & 596.7  & 579.3  & 627.7  & 3.3E-02 \\
freebase\_sampled    & 145,748,228 & 145,801,470 & 145,741,097 & 1.8E-04 & 2,409.9  & 2,424.5  & 2,493.8 & 1.5E-02 & 380.6    & 465.8    & 482.4   & 1.0E-01   \\
delicious-3d         & 139,996,876 & 139,993,400 & 139,999,596 & 1.8E-05 & 1,845.9  & 1,835.4  & 1,833.6 & 3.0E-03 & 290.3    & 350.6    & 328.5   & 7.7E-02   \\
NELL-1             & 145,175,629 & 145,155,883 & 145,155,081 & 6.5E-05 & 1,513.9 & 1,599.3 & 1,609.8 & 2.7E-02 & 256.0  & 300.6  & 292.8  & 6.9E-02 \\
\bottomrule
\end{tabular}%
}
\end{table*}

\subsubsection{Robustness and Feature Quality in Higher Orders}

To validate the generality of \gentensor{}, we performed a series of experiments on higher-order tensors with the aim of examining how well the generator preserves target structural properties as tensor dimensionality increases.
For this purpose, we generated synthetic tensors with orders $M = 6$, $7$, and $8$ where all mode sizes are fixed to 1000. 
Moreover, to demonstrate the robustness of the generator, we add perturbations in the input structural features (CV and densities) to show they do not drastically change the output structure.
Table~\ref{table:robust} summarizes the results under six settings, where each row corresponds to a different input configuration.



\begin{table}[!htb]
\centering
\caption{
Feature matching and structural robustness results, and average generation times for 6th–8th order tensors generated by \gentensor{} under six input settings.
In all settings, tensor sizes are set to 1000 for all modes.
Perturbed Settings 1, 2, 3, 4, and 5 respectively increase $cv_{fib}$, $cv_{nz}$, $d_{slc}$, $d_{fib}$, and $d_{nz}$ target values of “Reference Setting” by 10\%, with other features held fixed.
The values corresponding to perturbed features for each setting are shown in purple color.
The Feature Ratio is the ratio of resulting over target feature values. 
The Robustness Ratio is the ratio of the feature ratio in the perturbed setting over the feature ratio in the reference setting; and is reported only for the perturbed feature within each setting.
}
\label{table:robust}
\resizebox{\textwidth}{!}{%
\begin{tabular}{lcccccccccccccccccc}
\toprule
 &  & \multicolumn{5}{c}{Target Input Features} & \multicolumn{5}{c}{Resulting Generated   Features} & \multicolumn{5}{c}{Feature Ratio ( Resulting / Target )} &  &  \\
\cmidrule(lr){3-7}
    \cmidrule(lr){8-12}
     \cmidrule(lr){13-17}
\multirow{-2}{*}{\begin{tabular}[c]{@{}l@{}}Input   \\      Scenario\end{tabular}} & $M$ & $cv_{fib}$ & $cv_{nz}$ & $d_{slc}$ & $d_{fib}$ & $d_{nz}$ & $cv_{fib}$ & $cv_{nz}$ & $d_{slc}$ & $d_{fib}$ & $d_{nz}$ & $cv_{fib}$ & $cv_{nz}$ & $d_{slc}$ & $d_{fib}$ & $d_{nz}$ & \multirow{-2}{*}{\begin{tabular}[c]{@{}c@{}}Robust.\\      Ratio\end{tabular}} & \multirow{-2}{*}{\begin{tabular}[c]{@{}c@{}}Time \\      (sec)\end{tabular}} \\
\midrule
 & 6 & 1.0 & 1.0 & 1.0E-06 & 1.0E-08 & 1.0E-10 & 0.98 & 0.98 & 1.0E-06 & 9.9E-09 & 9.9E-11 & 0.98 & 0.98 & 1.00 & 0.99 & 0.99 & - & 1.43 \\
 & 7 & 1.0 & 1.0 & 1.0E-09 & 1.0E-11 & 1.0E-13 & 0.98 & 0.98 & 1.0E-09 & 9.9E-12 & 9.9E-14 & 0.98 & 0.98 & 1.00 & 0.99 & 0.99 & - & 1.43 \\
\multirow{-3}{*}{\begin{tabular}[c]{@{}l@{}}Reference\\      Setting\end{tabular}} & 8 & 1.0 & 1.0 & 1.0E-12 & 1.0E-14 & 1.0E-16 & 0.98 & 0.98 & 1.0E-12 & 9.9E-15 & 9.9E-17 & 0.98 & 0.98 & 1.00 & 0.99 & 0.99 & - & 1.46 \\
\midrule
 & 6 & {\color[HTML]{FC0CEB} \textbf{1.1}} & 1.0 & 1.0E-06 & 1.0E-08 & 1.0E-10 & {\color[HTML]{FC0CEB} \textbf{1.08}} & 0.98 & 1.0E-06 & 9.9E-09 & 9.9E-11 & {\color[HTML]{FC0CEB} \textbf{0.98}} & 0.98 & 1.00 & 0.99 & 0.99 & {\color[HTML]{FC0CEB} \textbf{0.999
}} & 1.42 \\
 & 7 & {\color[HTML]{FC0CEB} \textbf{1.1}} & 1.0 & 1.0E-09 & 1.0E-11 & 1.0E-13 & {\color[HTML]{FC0CEB} \textbf{1.08}} & 0.98 & 1.0E-09 & 9.9E-12 & 9.9E-14 & {\color[HTML]{FC0CEB} \textbf{0.98}} & 0.98 & 1.00 & 0.99 & 0.99 & {\color[HTML]{FC0CEB} \textbf{0.999
}} & 1.43 \\
\multirow{-3}{*}{\begin{tabular}[c]{@{}l@{}}Perturbed\\      Setting 1\\      ($cv_{fib} \times$1.1)\end{tabular}} & 8 & {\color[HTML]{FC0CEB} \textbf{1.1}} & 1.0 & 1.0E-12 & 1.0E-14 & 1.0E-16 & {\color[HTML]{FC0CEB} \textbf{1.08}} & 0.98 & 1.0E-12 & 9.9E-15 & 9.9E-17 & {\color[HTML]{FC0CEB} \textbf{0.98}} & 0.98 & 1.00 & 0.99 & 0.99 & {\color[HTML]{FC0CEB} \textbf{0.999
}} & 1.47 \\
\midrule
 & 6 & 1.0 & {\color[HTML]{FC0CEB} \textbf{1.1}} & 1.0E-06 & 1.0E-08 & 1.0E-10 & 0.98 & {\color[HTML]{FC0CEB} \textbf{1.08}} & 1.0E-06 & 9.9E-09 & 9.9E-11 & 0.98 & {\color[HTML]{FC0CEB} \textbf{0.98}} & 1.00 & 0.99 & 0.99 & {\color[HTML]{FC0CEB} \textbf{0.998
}} & 1.44 \\
 & 7 & 1.0 & {\color[HTML]{FC0CEB} \textbf{1.1}} & 1.0E-09 & 1.0E-11 & 1.0E-13 & 0.98 & {\color[HTML]{FC0CEB} \textbf{1.08}} & 1.0E-09 & 9.9E-12 & 9.9E-14 & 0.98 & {\color[HTML]{FC0CEB} \textbf{0.98}} & 1.00 & 0.99 & 0.99 & {\color[HTML]{FC0CEB} \textbf{0.998
}} & 1.46 \\
\multirow{-3}{*}{\begin{tabular}[c]{@{}l@{}}Perturbed\\      Setting 2\\      ($cv_{nz} \times$1.1)\end{tabular}} & 8 & 1.0 & {\color[HTML]{FC0CEB} \textbf{1.1}} & 1.0E-12 & 1.0E-14 & 1.0E-16 & 0.98 & {\color[HTML]{FC0CEB} \textbf{1.08}} & 1.0E-12 & 9.9E-15 & 9.9E-17 & 0.98 & {\color[HTML]{FC0CEB} \textbf{0.98}} & 1.00 & 0.99 & 0.99 & {\color[HTML]{FC0CEB} \textbf{0.998
}} & 1.48 \\
\midrule
 & 6 & 1.0 & 1.0 & {\color[HTML]{FC0CEB} \textbf{1.1E-06}} & 1.0E-08 & 1.0E-10 & 0.98 & 0.98 & {\color[HTML]{FC0CEB} \textbf{1.1E-06}} & 9.9E-09 & 9.9E-11 & 0.98 & 0.98 & {\color[HTML]{FC0CEB} \textbf{1.00}} & 0.99 & 0.99 & {\color[HTML]{FC0CEB} \textbf{1.000}} & 1.43 \\
 & 7 & 1.0 & 1.0 & {\color[HTML]{FC0CEB} \textbf{1.1E-09}} & 1.0E-11 & 1.0E-13 & 0.98 & 0.98 & {\color[HTML]{FC0CEB} \textbf{1.1E-09}} & 9.9E-12 & 9.9E-14 & 0.98 & 0.98 & {\color[HTML]{FC0CEB} \textbf{1.00}} & 0.99 & 0.99 & {\color[HTML]{FC0CEB} \textbf{1.000}} & 1.45 \\
\multirow{-3}{*}{\begin{tabular}[c]{@{}l@{}}Perturbed\\      Setting 3\\      ($d_{slc} \times$1.1)\end{tabular}} & 8 & 1.0 & 1.0 & {\color[HTML]{FC0CEB} \textbf{1.1E-12}} & 1.0E-14 & 1.0E-16 & 0.98 & 0.98 & {\color[HTML]{FC0CEB} \textbf{1.1E-12}} & 9.9E-15 & 9.9E-17 & 0.98 & 0.98 & {\color[HTML]{FC0CEB} \textbf{1.00}} & 0.99 & 0.99 & {\color[HTML]{FC0CEB} \textbf{1.000}} & 1.48 \\
\midrule
 & 6 & 1.0 & 1.0 & 1.0E-06 & {\color[HTML]{FC0CEB} \textbf{1.1E-08}} & 1.0E-10 & 1.03 & 0.98 & 1.0E-06 & {\color[HTML]{FC0CEB} \textbf{1.1E-08}} & 9.9E-11 & 1.03 & 0.98 & 1.00 & {\color[HTML]{FC0CEB} \textbf{0.99}} & 0.99 & {\color[HTML]{FC0CEB} \textbf{1.000}} & 1.48 \\
 & 7 & 1.0 & 1.0 & 1.0E-09 & {\color[HTML]{FC0CEB} \textbf{1.1E-11}} & 1.0E-13 & 1.03 & 0.98 & 1.0E-09 & {\color[HTML]{FC0CEB} \textbf{1.1E-11}} & 9.9E-14 & 1.03 & 0.98 & 1.00 & {\color[HTML]{FC0CEB} \textbf{0.99}} & 0.99 & {\color[HTML]{FC0CEB} \textbf{1.000}} & 1.49 \\
\multirow{-3}{*}{\begin{tabular}[c]{@{}l@{}}Perturbed\\      Setting 4\\      ($d_{fib} \times$1.1)\end{tabular}} & 8 & 1.0 & 1.0 & 1.0E-12 & {\color[HTML]{FC0CEB} \textbf{1.1E-14}} & 1.0E-16 & 1.03 & 0.98 & 1.0E-12 & {\color[HTML]{FC0CEB} \textbf{1.1E-14}} & 9.9E-17 & 1.03 & 0.98 & 1.00 & {\color[HTML]{FC0CEB} \textbf{0.99}} & 0.99 & {\color[HTML]{FC0CEB} \textbf{1.000}} & 1.54 \\
\midrule
 & 6 & 1.0 & 1.0 & 1.0E-06 & 1.0E-08 & {\color[HTML]{FC0CEB} \textbf{1.1E-10}} & 0.98 & 1.03 & 1.0E-06 & 9.9E-09 & {\color[HTML]{FC0CEB} \textbf{1.1E-10}} & 0.98 & 1.03 & 1.00 & 0.99 & {\color[HTML]{FC0CEB} \textbf{0.99}} & {\color[HTML]{FC0CEB} \textbf{0.999
}} & 1.50 \\
 & 7 & 1.0 & 1.0 & 1.0E-09 & 1.0E-11 & {\color[HTML]{FC0CEB} \textbf{1.1E-13}} & 0.98 & 1.03 & 1.0E-09 & 9.9E-12 & {\color[HTML]{FC0CEB} \textbf{1.1E-13}} & 0.98 & 1.03 & 1.00 & 0.99 & {\color[HTML]{FC0CEB} \textbf{0.99}} & {\color[HTML]{FC0CEB} \textbf{0.999
}} & 1.50 \\
\multirow{-3}{*}{\begin{tabular}[c]{@{}l@{}}Perturbed\\      Setting 5\\      ($d_{nz} \times$1.1)\end{tabular}} & 8 & 1.0 & 1.0 & 1.0E-12 & 1.0E-14 & {\color[HTML]{FC0CEB} \textbf{1.1E-16}} & 0.98 & 1.03 & 1.0E-12 & 9.9E-15 & {\color[HTML]{FC0CEB} \textbf{1.1E-16}} & 0.98 & 1.03 & 1.00 & 0.99 & {\color[HTML]{FC0CEB} \textbf{0.99}} & {\color[HTML]{FC0CEB} \textbf{0.999
}} & 1.58 \\
 \\
\bottomrule
\end{tabular}%
}
\end{table}

In our Reference Setting, the target CV parameters are set to $cv_{fib} = cv_{nz} = 1.0$; and the target densities are chosen so that the target setting yields approximately 1 million nonzero slices, 10 million nonzero fibers, and 100 million nonzeros for all cases with $M = 6$, $7$, and $8$.
For instance for  $M = 6$, it corresponds to assign $cv_{fib}$= $cv_{nz}$=$1.0$,  
$d_{slc}$=$10^{-6}$, $d_{fib}$=$10^{-8}$, and $d_{nz}$=$10^{-10}$ as input parameters.
In each “Perturbed Setting”, a single input feature of “Reference Setting” is increased by 10\%, with other features held fixed: 
Specifically, Perturbed Settings 1, 2, 3, 4, and 5 respectively increase the target values of $cv_{fib}$, $cv_{nz}$, $d_{slc}$, $d_{fib}$, and $d_{nz}$ by 10\%. 
For instance for $M = 6$, Perturbed Setting 1 changes $cv_{fib}$ to 1.1; Perturbed Setting 2 changes $cv_{nz}$ to 1.1; Perturbed Setting 3 updates $d_{slc}$ as $1.1 \times 10^{-6}$; Perturbed Setting 4 updates $d_{fib}$ as $1.1 \times 10^{-8}$; and Perturbed Setting 5 only updates $d_{nz}$ as $1.1 \times 10^{-10}$; while keeping other parameters the same as in the Reference Setting. 
For each setting, we compare the generated structural features against their target values.

For each setting, we compute the ratio between the resulting value over the target input value; to which we refer as the \textit{feature ratio}.
Results indicate that the generated tensors show close agreement with the specified input features across all orders. 
For example for the Reference Setting, the resulting CV and density values in the generated tensors deviate by less than 2\% and 1\% from their targets, respectively. 
Similarly in perturbed settings, \gentensor{} maintains highly accurate outputs. 
For all settings and tensor orders, the feature ratios remain between 0.98 and 1.03.
These results support that \gentensor{} can effectively reproduce desired structural characteristics even in higher-dimensional settings. 

To evaluate robustness, we compare the generator’s output in the perturbed setting to its output in the reference setting.  We define the \textit{robustness ratio} as the ratio of the feature ratio in the perturbed setting over the feature ratio in the reference setting. In other words, the robustness ratio measures how proportionally the output feature changes in response to a controlled change in the input feature. 
A robustness ratio close to 1.0 indicates that the generator reacts proportionally to the input perturbation, which is a desirable robustness property.
As shown in Table ~\ref{table:robust}, the robustness ratios range from 0.998 to 1.000 for all perturbed settings and tensor orders. 
These results demonstrate the structural robustness of \gentensor{}, since it
maintains consistent structural behavior under small deviations in input features.

A robustness ratio close to one indicates that the generator reacts proportionally to the input perturbation, which is a desirable robustness property. As shown in Table X, the robustness ratios across all perturbed features and tensor orders range from 0.998 to 1.000. These results demonstrate that the generator is robust, meaning it maintains consistent structural behavior under small deviations in input features.

The last column of Table~\ref{table:robust} reports the tensor generation time as an average of three runs. 
It is seen that perturbing CV or density does not significantly impact runtime. 
Furthermore, the runtime shows just a slight and expected increase as $M$ increases, indicating that dimensionality alone does not significantly impact the runtime of \gentensor{}.


\section{Discussion}
\label{sec:discussion}

\subsection{Related Work}
\label{sec:related-work}

To the best of our knowledge, there is only one study (and its extension) that extracts tensor features for optimizing sparse tensor computations in the literature: Sun et al.\citep{sun2020sptfs} proposed a framework, namely SpTFS, that automatically predicts the optimal storage format for CPD. 
SpTFS lowers the sparse tensor to fixed-sized matrices and gives them to convolutional neural networks (CNN) as inputs along with tensor features.
The authors improve SpTFS by adopting both supervised and unsupervised learning-based machine learning models in a recent work~\citep{sun2021input}.

The previous works \citep{sun2020sptfs, sun2021input} have considered the features for only one mode, assuming that the tensor is already sorted along that mode. However, real-world tensor sizes diverge significantly so that sizes in some modes reach millions while some are only orders of ten or even less. Therefore, considering the global values might result in losing some important information about the structure of the tensor.
Moreover, the feature extraction implementation of SpTFS is not publicly available in its official repository \citep{SpTFS}, which includes only the learning scripts and trained models. In contrast, our \featensor{} framework provides comprehensive multi-mode feature extraction without requiring any preprocessing. As a publicly available and parallelized tool, \featensor{} enables broader applicability and reproducibility, making it a more flexible and practical choice for large-scale sparse tensor analysis.
Other widely used libraries for tensor decomposition or sparse storage optimization, such as SPLATT \citep{smith2015splatt}, ParTI \citep{li2018parti}, and HiCOO \citep{li2018hicoo}, do not provide any documented standalone feature extraction module either. Thus, to the best of our knowledge, FeaTensor represents the first open-source framework offering parallel, multi-mode feature extraction for sparse tensors.

Although several sparse matrix and graph generators are proposed in the literature \citep{luszczek2020scalable, mpakos2023feature, kolda2014scalable, armstrong2008reinforcement, bonifati2020graph}, the studies on sparse tensor generators are very limited.
Chi and Kolda \citep{chi2012tensors} aim to produce low rank tensors; and Smith and Karypis \citep{smith2017accelerating} used their method to generate synthetic tensors whose values follow a Poisson distribution.
Baskaran et al. \citep{baskaran2012efficient} used MATLAB Tensor Toolkit \citep{kolda2006matlab} to generate synthetic sparse tensors but these are rather small tensors with less than one million nonzeros.
Due to the deficiency of publicly available sparse tensors, Sun et al.  \citep{sun2020sptfs} produced synthetic tensors by combining sparse matrices in Suite Sparse \citep{suitesparse} collection such that the elements of matrices form the modes of tensors. 
However real-world tensors are much sparser than real matrices and thus the structure of tensors generated by their method may differ significantly from the real ones.


\subsection{Limitations}
\label{sec:limit}

To the best of our knowledge, \featensor{} and \gentensor{} are the first open-source tools with feature-preserving tensor generation and efficient multi-method feature extraction for sparse tensors.
Despite their demonstrated utility, some limitations remain that offer opportunities for future enhancement.
While this work primarily focuses on replicating structural sparsity patterns, it does not explicitly incorporate constraints on nonzero values or tensor rank. 
These aspects may affect downstream tasks such as classification, regression, or clustering, where the semantics of nonzero entries are important.

Additionally, although both \featensor{} and \gentensor{} are parallelized via OpenMP for shared-memory systems, their applicability to distributed-memory environments remains unexplored. Extending these tools to scale across nodes would broaden their utility in HPC settings.
Addressing these aspects would further enhance the quality of these tools, making it a promising direction for future research.

\gentensor{} is designed with the flexibility to incorporate any distribution. However, its current version utilizes normal and log-normal distributions to determine the nonzero layout. Furthermore, the positions of the nonzeros, as well as the indices of nonzero slices and fibers, are selected uniformly, and nonzero values are drawn from a uniform distribution.
As part of future work, we plan to extend \gentensor{} by exploring alternative distribution models for both layout and value generation.

In this work, the effectiveness of \gentensor{} has mainly been demonstrated through matching structural features and replicating CPD performance. These evaluation metrics were selected to capture both the structural fidelity and practical applicability of the generated tensors.
Nonetheless, we acknowledge that other potential indicators could offer complementary perspectives. For instance, tensor rank or spectral characteristics might provide insight into latent structure fidelity. Moreover, evaluating the generated tensors under alternative decomposition models, such as Tucker, could reveal behavioral differences not captured by CPD alone. Graph-based similarity metrics also present a potential direction. 
While such alternatives may broaden the scope of evaluation, they often depend on specific applications or assumptions, which our current generic generator intentionally avoids. 
Exploring these dimensions in a targeted context remains a promising direction for future work.

\subsection{Conclusion}
\label{sec:conclusion}

In this study, we introduce two tools, \featensor{} and \gentensor{}, which we designed and developed to advance research in sparse tensor operations.
\featensor{} is a feature extraction framework for sparse tensors that provides four different methods, prioritizing efficiency in extracting tensor features.
It serves as the first publicly available, multi-mode, and parallel feature extraction framework for sparse tensors.
This contribution is particularly valuable, as feature extraction itself is challenging and computationally expensive due to the large number of fibers and slices in real sparse tensors. We evaluate the performance of various feature extraction methods and observe that the methods introduced in this work outperform conventional approaches.

\gentensor{} is a smart sparse tensor generator that adheres to a comprehensive set of tensor features. Experimental results validate its effectiveness in mimicking real tensors, both in terms of sparsity patterns and tensor decomposition performance. A key advantage of \gentensor{} is its use of size-independent features, enabling the generation of tensors at different scales while preserving essential properties of real tensors. This capability facilitates the creation of large synthetic sparse tensor datasets that exhibit characteristics and behavior similar to real-world data.
This is particularly useful for performance and scaling experiments involving tensor decomposition kernels that depend on tensor sparsity patterns. 
Moreover, experimental results demonstrate that \gentensor{} generalizes well to higher-order tensors and is robust to typical variations in the input feature set. The generator can therefore be reliably used in realistic benchmarking and modeling scenarios where input characteristics may not be precisely known.


\section*{Funding}

This work was supported by the European Research Council (ERC) under the European Union’s Horizon 2020 research and innovation programme (grant agreement No 949587) and EuroHPC Joint Undertaking through the grant agreement No 956213.


\section*{Acknowledgments}

We thank Eren Yenigul from Koç University for his contributions in developing the \featensor{} tool.

\bibliographystyle{Frontiers-Vancouver} 
\bibliography{references}



\end{document}